\documentclass[useAMS,usenatbib,usegraphicx]{mn2e}
\usepackage{epsfig}
\usepackage{aas_macros} % required to get journal abbreviations
\usepackage{amsmath}
\usepackage{amssymb}
\usepackage{appendix}
\usepackage{booktabs}
\usepackage{upgreek}
\usepackage{url}
\usepackage[usenames]{color}

\newcommand{\ion}[2]{{#1}\,{\sc #2}}
\newcommand{\specline}[3]{{#1}\,{\sc #2}\:{#3}}
\newcommand{\kepler}{\textit{Kepler}}
\usepackage{graphicx}
\usepackage{rotating}
\usepackage{color}
\usepackage{pifont}
\usepackage{longtable}
\usepackage{pdflscape,lscape}
\usepackage{natbib}

\title[Spectroscopic survey of \textit{Kepler} stars]{Spectroscopic survey of \textit{Kepler} stars.\\
II. FIES/NOT observations of A- and F-type stars}
\author[E.\,Niemczura et al.]{E.\,Niemczura$^1$\thanks{E-mail: eniem@astro.uni.wroc.pl},
M.\,Poli\'{n}ska$^2$,
S.\,J.\,Murphy$^{3,4}$,
B.\,Smalley$^5$,
Z.\,Ko\l{}aczkowski$^1$,\and
J.\,Jessen-Hansen$^6$,
K.\,Uytterhoeven$^{7,8}$,
J.\,M.\,Lykke$^9$,
A.\,Trivi\~{n}o Hage$^{7,8}$,
G.\,Michalska$^1$
\\
$^1$ Instytut Astronomiczny, Uniwersytet Wroc{\l}awski, Kopernika 11, 51-622 Wroc{\l}aw, Poland\\
$^2$ Astronomical Observatory Institute, Faculty of Physics, A. Mickiewicz University, S\l{}oneczna 36, 60-286 Pozna\'{n}, Poland \\
$^3$ Sydney Institute for Astronomy (SIfA), School of Physics, University of Sydney NSW 2006, Australia\\
$^4$ Stellar Astrophysics Centre, Department of Physics and Astronomy, Aarhus University, 8000 Aarhus C, Denmark\\
$^5$ Astrophysics Group, Keele University, Staffordshire, ST5 5BG, United Kingdom\\
$^6$ Stellar Astrophysics Centre, Department of Physics and Astronomy, Aarhus University, Ny Munkegade 120, DK-8000 Aarhus C, Denmark\\
$^7$ Instituto de Astrofisica de Canarias, E-38205 La Laguna, Tenerife, Spain\\
$^8$ Universidad de La Laguna, Departamento de Astrofisica, E-38206 La Laguna, Tenerife, Spain\\
$^9$ Centre for Star and Planet Formation, Niels Bohr Institute \& Natural History Museum of Denmark, University of Copenhagen, \\ 
$\ $ \O{}ster Voldgade 5-7, 1350, Copenhagen K., Denmark}

\begin{document}
\def\teff{${T}_{\rm eff}$}
\def\kms{{km\,s}$^{-1}$}
\def\logg{$\log g$}
\def\ksi{$\xi_{\rm t}$}
\def\micro{$\xi_{\rm t}$}
\def\macro{$\zeta_{\rm RT}$}
\def\rad{$v_{\rm r}$}
\def\vsini{$v\sin i$}
\def\ebv{$E(B-V)$}
\def\kepler{\textit{Kepler}}

\date{Accepted ... Received ...; in original form ...}

\pagerange{\pageref{firstpage}--\pageref{lastpage}} \pubyear{2002}

\maketitle

\label{firstpage}

\begin{abstract}
We have analysed high-resolution spectra of $28$ A and $22$ F stars in the \kepler\ field, observed with the FIES spectrograph at the Nordic Optical Telescope. 
We provide spectral types, atmospheric parameters and chemical abundances for $50$ stars. Balmer, \ion{Fe}{i}, and \ion{Fe}{ii} lines were used 
to derive effective temperatures, surface gravities, and microturbulent velocities. We determined chemical abundances and projected rotational velocities using a spectrum synthesis technique. 
Effective temperatures calculated by spectral energy distribution fitting are in good agreement with those determined from the spectral line analysis. 
The stars analysed include chemically peculiar stars of the Am and $\lambda$\,Boo types, as well as stars with approximately solar chemical abundances. 
The wide distribution of projected rotational velocity, \vsini, is typical for A and F stars.
The microturbulence velocities obtained are typical for stars in the observed temperature and surface gravity ranges. 
Moreover, we affirm the results of Niemczura et al., that Am stars do not have systematically higher microturbulent velocities than normal stars of the same temperature.

\end{abstract}

\begin{keywords}
stars: abundances -- stars: chemically peculiar -- stars: rotation -- stars: general.
\end{keywords}

%&&&&&&&&&&&&&&&&&&&&&&&&&&&&&&&&&&&&&&&&&&&&&&&&&&&&&&&&&&&&&&&&&&&&&&&&&&&&&&&&&&&&&&&&&&&&&&&&&&&&&&&&&&&&&&&&&&

\section{Introduction}
\label{sec:introduction}

The high-precision \kepler\ photometric time series are excellent for studying the variability of stars across the Hertzsprung-Russell (H-R) diagram \citep{Kjeldsen, 2015AJ....150..133G}. 
The ultra-high precision and long duration offer unparalleled opportunities to discover and characterize stellar pulsations 
and consecuently to make progress in modelling stellar evolution and internal structure using asteroseismology.

To perform an asteroseismic study, the precise pulsation frequencies, mode identification, and atmospheric parameters (effective temperature $T_{\rm eff}$, 
surface gravity $\log g$, chemical abundances, rotational velocity $v \sin i$) are the necessary ingredients. Pulsation frequencies can be determined from the analysis 
of the precise \kepler\ photometry. The best way to obtain information about atmospheric parameters is the investigation of high-resolution spectra. For this reason, 
ground-based spectroscopic observations of stars in the \kepler\ field began before the mission \citep{Uytterhoeven2010}.

High- and medium-resolution spectra have been collected and analysed for a substantial number of A and F-type stars 
\citep[e.g.][hereafter Paper\,I]{2011MNRAS.411.1167C, 2012MNRAS.422.2960T, 2013A&A...556A..52T, 2013MNRAS.431.3685T, 2015MNRAS.450.2764N}. 
The atmospheric parameters obtained in those investigations are of great importance for seismic modelling of $\delta$\,Scuti, $\gamma$\,Doradus, and $\gamma$\,Dor/$\delta$\,Sct 
hybrid stars, which occupy this region of the H-R diagram. The hybrid variables are of particular interest, showing g-mode as well as p-mode pulsations. 
In such cases, two regions of the stellar interior can be probed because the g-modes propagate in the layers between the convective core and the envelope while the 
p-modes probe the outer layers of the envelope \citep{2015arXiv150200175G}.

An additional benefit of studying atmospheric parameters of pulsating A and F stars is 
the possibility to observationally define instability regions of $\delta$\,Scuti, $\gamma$\,Doradus, and $\gamma$\,Dor/$\delta$\,Sct hybrids on the H-R diagram. 
The question of purity of the pulsational instability strips has been addressed with high-resolution spectroscopy once before, when the improvement over SDSS 
photometry showed the DAV instability strip to be pure \citep{2015MNRAS.447.3948M}.

From the analysis of \kepler\ data, \citet{Grigahcene2010}, and \citet{2011A&A...534A.125U} discovered a large number of hybrid star candidates,
occupying a region of the $T_{\rm eff}$-$\log g$ diagram broader than predicted by the current stellar pulsation theory \citep[e.g.][]{2015arXiv150200175G}.
However, the correct interpretation of the observed low-frequency modes can be difficult. 
For instance, \citet{balonadziembowski} analysed over $12,000$ \kepler\ stars and identified $1568$ candidate $\delta$\,Sct pulsators. 
They defined a group of $\delta$\,Sct stars with low-frequency modes similar to those characteristic for $\gamma$\,Dor stars. 
Photometric atmospheric parameters place all of these stars in the $\gamma$\,Dor region of the H-R diagram.
Accurate stellar atmospheric parameters are obligatory to place the observed $\gamma$\,Dor/$\delta$\,Sct hybrids in the H-R diagram.

The investigation of \kepler\ data revealed A and F-type stars located in the $\delta$\,Scuti or $\gamma$\,Doradus instability regions but showing 
no variability \citep[][]{2015arXiv150200175G, 2015MNRAS.447.3948M}. The simplest explanation of this phenomenon is that the atmospheric parameters are incorrect.
If so, the determination of accurate values would place the star outside the instability regions. Other explanations are observational 
constraints preventing the detection of oscillations with amplitudes below the noise level of the data, or the existence of a physical mechanism that inhibits pulsations for some stars. 
The problem of non-pulsating A and F stars is widely discussed in \citet[][]{2015arXiv150200175G} and \citet[][]{2015MNRAS.447.3948M}.

The analysis of large samples of A and F stars helps in finding answers to many questions concerning these stars and their atmospheres in general, 
like the \vsini\ distribution and the influence of fast rotation on chemical abundances, the dependence of microturbulent velocity on effective temperature and surface gravity,
and the reasons for peculiar abundances of chemical elements among A and F stars.

This paper is organised as follows. In Sect.\,\ref{sec:observations}, we describe observational data used in the analysis, methods of data reduction, calibration and normalisation. 
The methods applied for spectral classification and for determining the atmospheric parameters of the investigated objects,
together with the sources of possible errors are given in Sect.\,\ref{sec:methods}. 
The results are discussed in Sect.\,\ref{sec:results}. The sample of chemically peculiar (CP) stars are discussed in Sect.\,\ref{sec:abundances}. 
Positions of the stars in the $\log$\,\teff\ -- \logg\ diagram and the discussion of their pulsation properties are shown in Sect.\,\ref{sec:hr}. 
Conclusions and future prospects are given in Sect.\,\ref{sec:conclusions}.

\begin{table*}
\centering
\caption{Journal of spectroscopic observations, the derived spectral classes, and variability classification. 
$N$ is the number of available spectra; SpT1 indicates the spectral classification from the literature, whereas SpT2 means spectral classifications 
obtained in this work. Additional notations for luminosity class: 
 ``b'' -- lower luminosity main-sequence star; 
 ``a'' or ``a+'' -- higher luminosity main-sequence star; 
 ``IV-V'' -- between IV and V; 
 ``IV/V'' -- either IV or V; 
 ``n'' -- nebulous (e.g. broad-lined); 
 ``s'' -- sharp lines; 
 ``nn'' -- very rapid rotators; 
 ``met wk'' -- weak metal lines; 
 ``met str'' -- strong metal lines. 
In the last column additional information taken from spectra and \kepler\ data are presented. 
``SB2'' means double-lined spectroscopic binary; 
``EB'' -- eclipsing binary; 
``ell. var.'' -- ellipsoidal variable; 
``hybrid'' -- frequencies of $\gamma$\,Dor and $\delta$\,Sct visible in periodogram;
``HADS'' -- high amplitude $\delta$\,Sct variable;
``cont.'' -- \kepler\ data contaminated by a nearby star; 
``unknown'' -- unknown variability type.} 
\label{table1}
\begin{scriptsize}

\begin{tabular}{llccrlll}
\toprule
KIC                & Obs. & N &Resolving   & $V$  & SpT1                   & SpT2                     & Notes \\
Number             & year &   &power       &[mag] &(literature)            & (this work)              &       \\
\midrule
KIC\,2162283       & 2010 & 1 &med &  9.47 & F2$^1$                        & Am+F                                  & EB, ell. var., $\delta$\,Sct                                    \\ 
KIC\,2694337       & 2010 & 1 &low & 10.35 & F0.5\,IVs$^6$                 & A9 (met str F3)                       & $\gamma$\,Dor/$\delta$\,Sct hybrid                              \\
KIC\,3217554       & 2010 & 1 &med &  9.56 & A5$^1$                        & A3\,IVn                               & SB2, $\gamma$\,Dor ($\gamma$\,Dor/$\delta$\,Sct hybrid?), cont. \\
KIC\,3219256       & 2010 & 1 &med &  8.31 & A8\,V$^2$                     & A9\,V                                 & $\gamma$\,Dor/$\delta$\,Sc hybrid                               \\ %HD\,178306
KIC\,3331147       & 2011 & 1 &med & 10.06 & --                            & F0.5\,Vs                              & $\gamma$\,Dor                                                   \\
KIC\,3429637       & 2010 & 2 &med &  7.72 & kF2\,hA9\,mF3$^3$,F0\,III$^2$ & kF0\,hA8\,mF0\,IIIas                  & $\delta$\,Sct ($\gamma$\,Dor/$\delta$\,Sct hybrid?)             \\ %HD\,178875
KIC\,3453494       & 2010 & 1 &med &  9.53 & A7\,IV$^4$                    & A4\,Vn                                & $\gamma$\,Dor/$\delta$\,Sct hybrid \\
KIC\,3858884       & 2010 & 2 &med &  9.28 & F5$^5$                        & kA8\,hF0\,mF2.5\,III Am:              & SB2, EB, $\delta$\,Sct \\
KIC\,3868420       & 2011 & 1 &med &  9.88 & --                            & A9\,Vs (Ca wk A4)                     & HADS ($\gamma$\,Dor/$\delta$\,Sct hybrid?) \\
KIC\,4647763       & 2012 & 2 &low & 10.89 & --                            & F0.5\,IV-III                          & $\delta$\,Sct/$\gamma$\,Dor hybrid\\
KIC\,4840675       & 2010 & 1 &med &  9.66 & F0\,Vn$^{13}$                 & F0\,Vn                                & SB2, $\delta$\,Sct/$\gamma$\,Dor hybrid\\
KIC\,5641711       & 2010 & 2 &low & 10.51 & F0$^1$                        & F0\,Vn                                & SB2, $\delta$\,Sct/$\gamma$\,Dor hybrid \\ %HD\,226029
KIC\,5880360       & 2010 & 1 &med &  8.75 & A0$^1$                        & A3\,IVan                              & unknown ($\gamma$\,Dor?), $0-7$\,c/d \\ %HD\,184380
KIC\,5988140       & 2010 & 1 &med &  8.81 & A8\,IIIs$^6$                  & A9\,IV                                & $\delta$\,Sct \\ %HD\,188774
KIC\,6123324       & 2010 & 1 &med &  8.71 & F0$^1$                        & F2.5\,Vn                              & $\delta$\,Sct/$\gamma$\,Dor hybrid? \\ %HD\,183281
KIC\,6279848       & 2010 & 1 &med &  9.23 & F0$^1$                        & F1.5\,IVs (met wk A9)                 & $\delta$\,Sct/$\gamma$\,Dor hybrid \\ %HD\,181877
KIC\,6370489       & 2011 & 1 &med &  8.79 & F8\,V$^7$                     & F5.5\,V                               & solar-like oscillations \\
KIC\,6443122       & 2011 & 1 &med & 11.10 & --                            & F0\,IVs                               & $\delta$\,Sct/$\gamma$\,Dor hybrid \\
KIC\,6670742       & 2010 & 1 &med &  9.10 & A5$^1$                        & F0\,Vnn (met wk A7)                   & $\delta$\,Sct/$\gamma$\,Dor hybrid \\
KIC\,7119530       & 2010 & 1 &med &  8.44 & A3\,IVn$^6$                   & A3\,IVn                               & $\delta$\,Sct/$\gamma$\,Dor hybrid \\ %HD\,183787
KIC\,7122746       & 2012 & 2 &low & 10.62 & --                            & A5\,Vn                                & $\delta$\,Sct/$\gamma$\,Dor hybrid? \\
KIC\,7668791       & 2010 & 1 &med &  9.21 & A2$^1$                        & A5/7\,IV/V                            & $\delta$\,Sct \\ %HD\,178120
KIC\,7748238       & 2011 & 1 &med &  9.54 & A9.5\,V-IV$^4$                & F1\,V (met wk F0)                     & $\delta$\,Sct/$\gamma$\,Dor hybrid \\%HD\,181985 
KIC\,7767565       & 2010 & 1 &med &  9.29 & kA5\,hA7\,mF1\,IV Am$^6$      & kA5\,hA7\,mF1\,IV Am                  & $\gamma$\,Dor \\ %HD\,186995 
KIC\,7773133       & 2012 & 1 &low & 10.93 & --                            & F0\,IIIas                             & $\delta$\,Sct ($\delta$\,Sct/$\gamma$\,Dor hybrid?)\\
KIC\,7798339       & 2010 & 1 &med &  7.85 & F3\,IV$^2$                    & F1.5\,IV                              & $\gamma$\,Dor \\%HD\,173109 
KIC\,8211500       & 2010 & 1 &med &  8.08 & A5\,IV$^8$                    & A6\,V                                 & unknown ($\gamma$\,Dor?)\\%HD\,173978 
KIC\,8222685       & 2010 & 1 &med &  8.89 & F0\,V$^9$                     & F0\,V                                 & $\gamma$\,Dor \\%HD\,179336 
KIC\,8694723       & 2011 & 1 &med &  8.92 & G0\,IV$^7$                    & hF6\,mF2\,gF5\,V                      & solar-like oscillations \\
KIC\,8827821       & 2012 & 1 &low & 11.14 & --                            & A8\,IV-V                              & $\delta$\,Sct/$\gamma$\,Dor hybrid? \\
KIC\,8881697       & 2010 & 1 &low & 10.50 & A5$^9$                        & A8\,Vn                                & $\delta$\,Sct/$\gamma$\,Dor hybrid \\
KIC\,8975515       & 2010 & 1 &med &  9.48 & A6\,V:$^6$                    & A7\,IV                                & SB2, $\delta$\,Sct/$\gamma$\,Dor hybrid \\%HD\,188538
KIC\,9229318       & 2010 & 1 &med &  9.59 & F0$^1$                        & A8\,III (met str F0)                  & $\delta$\,Sct/$\gamma$\,Dor hybrid \\
KIC\,9349245       & 2010 & 1 &med &  8.11 & Am:$^{10}$                    & F2\,IIIs                              & constant \\ %HD\,185658
KIC\,9408694       & 2011 & 2 &med & 11.48 & --                            & F0\,IVs + ?                           & HADS ($\delta$\,Sct/$\gamma$\,Dor hybrid?)\\
KIC\,9410862       & 2012 & 2 &low & 10.78 & --                            & F6\,V                                 & solar-like oscillations \\ 
KIC\,9650390       & 2010 & 1 &med &  9.32 & A0$^1$                        & A3\,IV-Vnn (met wk A2.5)              & $\delta$\,Sct/$\gamma$\,Dor hybrid \\
KIC\,9655114       & 2010 & 1 &low & 12.03 & A4$^{11}$                     & A5\,Vn                                & SB2, $\delta$\,Sct + cont. \\
KIC\,9656348       & 2010 & 3 &low & 10.28 & --                            & hA9\,kA4m\,A4\,V  $\lambda$\,Boo      & $\delta$\,Sct/$\gamma$\,Dor hybrid \\
KIC\,9828226       & 2012 & 1 &low &  9.92 & hA2\,kA0\,mB9\,Vb$^6$         & hA2\,kB9.5\,mB9\,Vb $\lambda$\,Boo    & $\delta$\,Sct/$\gamma$\,Dor hybrid \\
KIC\,9845907       & 2012 & 3 &low & 11.42 & --                            & A4\,IVs                               & $\delta$\,Sct \\
KIC\,9941662       & 2011 & 1 &med &  9.95 & A0$^1$                        & kA2\,hA5\,mA7\,(IV) Am:               & constant, planet transit \\ 
KIC\,9970568       & 2010 & 1 &med &  9.58 & A4\,IVn$^6$                   & A8\,Vnn (met wk A4)                   & $\delta$\,Sct/$\gamma$\,Dor hybrid\\ %HD\,188832 
KIC\,10030943      & 2011 & 2 &med & 11.45 & --                            & F1.5\,IV-V                            & ell. var., $\delta$\,Sct \\
KIC\,10341072      & 2010 & 1 &med &  8.95 & F8$^1$                        & F6\,III                               & unknown \\ %HD\,183954
KIC\,10661783      & 2011 & 1 &med &  9.53 & A2$^1$                        & A5\,IV                                & EB, $\delta$\,Sct \\
KIC\,10717871      & 2012 & 1 &low & 10.52 & --                            & F0\,V                                 & $\gamma$\,Dor/$\delta$\,Sct hybrid\\
KIC\,11044547      & 2012 & 1 &low &  9.94 & A0$^1$                        & A2\,Vs                                & $\delta$\,Sct ($\delta$\,Sct/$\gamma$\,Dor hybrid?)\\
KIC\,11622328      & 2010 & 1 &med &  9.42 & A2$^1$                        & A4.5\,Vn                              & unknown ($\gamma$\,Dor?) \\
KIC\,11973705      & 2010 & 1 &med &  9.09 & B8.5IV-V$^{12}$               & hF0.5\,kA2.5m\,A2.5\,V $\lambda$\,Boo & ell. var., $\delta$\,Sct/$\gamma$\,Dor hybrid \\ 

\bottomrule
\end{tabular}

{$^1$ } \citet{1975AGK3..C......0H}; 
{$^2$ } \citet{2011MNRAS.411.1167C}; 
{$^3$ } \citet{1984ApJ...285..247A}; 
{$^4$ } \citet{2012MNRAS.422.2960T}; 
{$^5$ } \citet{2001KFNT...17..409K}; 
{$^6$ } Paper\,I; 
{$^7$ } \citet{2013MNRAS.434.1422M}; 
{$^8$ } \citet{2015MNRAS.447.3948M}; 
{$^9$ } \citet{1952ApJ...116..592M}; 
{$^{10}$ } \citet{2015MNRAS.451..184C}; 
{$^{11}$ } \citet{1972A&A....16..315L};
{$^{12}$ } \citet{2011A&A...526A.124L}.
{$^{13}$ } this work

\end{scriptsize}
\end{table*}
%&&&&&&&&&&&&&&&&&&&&&&&&&&&&&&&&&&&&&&&&&&&&&&&&&&&&&&&&&&&&&&&&&&&&&&&&&&&&&&&&&&&&&&&&&&&&&&&&&&&&&&&&&&&&&&&&&&

\section{Spectroscopic surveys of \kepler\ stars}
\label{sec:observations}

In this work we analyse the spectra taken with the cross-dispersed, fibre-fed \'{e}chelle spectrograph FIES \citep{2014AN....335...41T},
attached to the $2.56$-m Nordic Optical Telescope (NOT) located on La Palma (Canary Islands, Spain).
The obtained spectra have `medium' $R$\,$\sim$\,$46\,000$ or `low' \mbox{$R$\,$\sim$\,$25\,000$} resolving power
and cover the visual spectral range from $3700$ to $7300$\,{\AA}. Signal-to-noise (S/N) ratio for an individual spectrum at $5500$\,{\AA} is $80$--$100$. 

The spectra have been reduced with a dedicated Python software package FIEStool\footnote{\url{http://www.not.iac.es/instruments/fies/fiestool/FIEStool.html}} \citep{2014AN....335...41T}
based on {\sc iraf}\footnote{Image Reduction and Analysis Facility, \url{http://iraf.noao.edu/}}.
The standard procedures have been applied, which includes bias subtraction, extraction of scattered light produced by the optical system, cosmic ray filtering,
division by a normalized flat-field, wavelength calibration by a ThArNe lamp, and order merging.
Normalisation to the continuum was performed using the standard {\sc iraf} procedure {\it continuum}.
The accuracy of the normalisation was tested by fitting synthetic spectra to four Balmer lines (H$\alpha$, H$\beta$, H$\gamma$, H$\delta$) in the spectrum of each star.
The normalisation process was successful if all hydrogen lines can be fitted satisfactorily with the same \teff.
Otherwise, the whole procedure is repeated. If several spectra were available for a star, we analysed the averaged spectrum.
The averaging process is applied after normalization and forms an additional test of the quality of the normalization process.

The data analysed here were collected during the spectroscopic survey carried out between $2010$ and $2012$.
The stars were selected on the basis of spectral types and photometric atmospheric parameters available
in the \kepler\ Input Catalogue \citep[KIC, ][]{brown} as a potential $\delta$\,Sct, $\gamma$\,Dor or hybrid variables.

Table\,\ref{table1} lists the stars analysed and gives their KIC\ numbers, year of the observations, number of spectra and their resolution, visual ($V$) magnitudes,
and spectral types taken from the literature and determined in the current work.
We also provide notes on chemical peculiarity resulting from spectral classification and spectroscopic analysis,
type of binarity, and pulsation characteristics obtained from \kepler\ data analysis. 

%&&&&&&&&&&&&&&&&&&&&&&&&&&&&&&&&&&&&&&&&&&&&&&&&&&&&&&&&&&&&&&&&&&&&&&&&&&&&&&&&&&&&&&&&&&&&&&&&&&&&&&&&&&&&&&&&&&

\section{Atmospheric parameters}
\label{sec:methods}

The basic information about the stars and the initial stellar atmospheric parameters were derived from spectral classification, spectral energy distribution (SED) 
analysis, and taken from the catalogue of \citet[][hereafter H2014]{huber2014}. Effective temperatures, surface gravities and metallicities determined by these methods were 
used as input parameters for the high-resolution spectroscopic data analysis. The atmospheric parameters are refined by analysis of the spectral lines, guided by our spectral classifications.

\begin{itemize}
\item[(i)] {\it Spectral classification} -- We compared the spectra of the studied stars with those of standards \citep{gray&corbally2009}.
Our sample consists of A and early F stars, so the spectral type is derived from hydrogen, \ion{Ca}\,K and metal lines.
For non-CP stars these three sets of lines provide the same spectral type. 
For CP stars (Am, marginal Am:, $\lambda$\,Bo\"{o}tis) different spectral types are derived. 
For early A stars the luminosity class is obtained from hydrogen lines, whereas for later types, the lines of ionised \ion{Fe} and \ion{Ti} became useful.

The literature and new spectral classifications of the investigated stars are given in Table~\ref{table1}.
The spectral classification process revealed CP stars of Am, Am: and $\lambda$\,Bo\"{o}tis type. 
Some of these stars were already known peculiar stars \citep[KIC\,3429637: Am:,][]{murphyetal2012},
\citep[KIC\,7767565 and KIC\,9828226: Am and $\lambda$\,Boo,][]{2015MNRAS.450.2764N}. 
We classified one new Am: star (KIC\,3858884) and two $\lambda$\,Boo objects (KIC\,9656348, KIC\,11973705).

Eight stars from our sample were also analysed in Paper\,I based on the high-resolution HERMES spectra (see Table\,\ref{table1}). 
We took this opportunity to check the influence of spectral resolution, S/N and normalisation process on the spectral classification results. 
No differences in classification were found for KIC\,7119530 (A3\,IVn) and CP stars KIC\,7767565 (kA5\,hA7\,mF1\,IV Am),
KIC\,9828226 (hA2\,kA0\,mB9\,Vb and hA2\,kB9.5\,mB9\,Vb, the negligible difference in spectral type from Ca lines).
Minor differences, none more than one temperature subclass or one luminosity class were found for KIC\,5988140 
(A8\,IIIs and A9\,IV), KIC\,8975515 (A6\,V: and A7\,IV), and KIC\,2694337 (F0.5\,IVs and A9\,met strong F3). 
Larger differences were found for the very fast rotating star KIC\,9970568 (A4\,IVn and A8\,Vnn met wk A4).
This could be a consequence of differences in the normalisation of the broad, blended spectral lines.

%&&&&&&&&&&&&&&&&&&&&&&&&&&&&&&&&&&&&&&&&&&&&&&&&&&&&&&&&&&&&&&&&&&&&&&&&&&&&&&&&&&&&&&&&&&&&&&&&&&&&&&&&&&&&&&&&&&

\begin{figure*}
\centering
\includegraphics[width=18cm,angle=0]{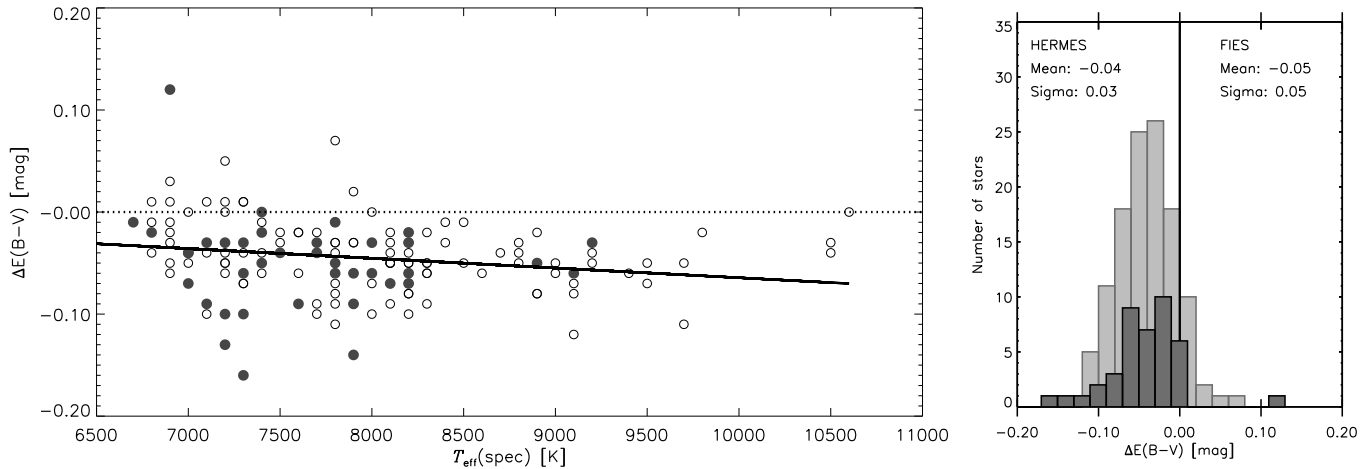}
\caption[]{Left panel: the differences $\Delta E(B-V) = E(B-V){\rm(Na)} - E(B-V){\rm(KIC)}$ as a function of \teff({\sc Fe}).
A linear function is fitted (solid line). The open circles represent results taken from Paper\,I.
The filled circles indicate the results obtained here. Right panel: the distribution of differences $\Delta E(B-V)$ for Paper\,I stars (HERMES observations, light gray histogram)
and objects analysed here (FIES observations, dark gray histogram). Arithmetic means and standard deviations of both distributions are indicated in the figure.}
\label{ebvcomparison}
\end{figure*}

%&&&&&&&&&&&&&&&&&&&&&&&&&&&&&&&&&&&&&&&&&&&&&&&&&&&&&&&&&&&&&&&&&&&&&&&&&&&&&&&&&&&&&&&&&&&&&&&&&&&&&&&&&&&&&&&&&&

\item[(ii)] {\it Interstellar reddening} -- SEDs are affected by interstellar reddening \ebv.
Therefore, we estimated this parameter using the equivalent width of the interstellar Na\,D$_2$ (5889.95\,{\AA}) line. 
\ebv\ values were obtained from the relation given by \citet{1997A&A...318..269M}.
The equivalent widths of the individual components were measured for those with resolved multi-component interstellar Na\,D$_2$ lines.
Because the interstellar reddening is additive \citep{1997A&A...318..269M}, the total \ebv\ is the sum of the reddening per component.

In Fig.\,\ref{ebvcomparison}, the spectroscopic interstellar reddenings $E(B-V){\rm (Na)}$ were compared with \ebv\ values from the KIC catalogue, $E(B-V){\rm (KIC)}$,
for stars analysed here and those taken from Paper\,I. The current results fully confirm those obtained in Paper\,I.
The photometric interstellar reddenings given in the KIC are higher than $E(B-V){\rm (Na)}$ and the differences increase with effective temperature.
The highest differences between $E(B-V){\rm (Na)}$ and $E(B-V){\rm (KIC)}$ were obtained for faint stars (KIC\,10030943, KIC\,6443122, KIC\,9408694, KIC\,9845907).\\

%&&&&&&&&&&&&&&&&&&&&&&&&&&&&&&&&&&&&&&&&&&&&&&&&&&&&&&&&&&&&&&&&&&&&&&&&&&&&&&&&&&&&&&&&&&&&&&&&&&&&&&&&&&&&&&&&&&

\item[(iii)] {\it Spectral Energy Distributions} -- Stellar effective temperatures \teff\ were determined from SEDs,
constructed from the available photometry and TD-1 ultraviolet fluxes \citep{1979BICDS..17...78C}.
For photometric data we searched the same photometric catalogues as in Paper\,I. 
The constructed SEDs were de-reddened using the appropriate analytical extinction fits of \citet{1979MNRAS.187P..73S} and \citet{1983MNRAS.203..301H} for the ultraviolet
and the optical and infrared spectral ranges, respectively.

As in Paper~I, we obtained \teff\ values using a non-linear least-squares fit to the SED. We fixed $\log g = 4.0$ and [M/H]\,$=0.0$ for all of the stars, since they are poorly constrained by the SED. The results are given in Table\,\ref{parameters1}. The uncertainties in \teff\ include the formal least-squares error and those due to the assumed uncertainties in $E(B-V)$ of $\pm0.02$\,mag, \logg\ of $\pm0.5$ and [M/H] of $\pm0.5$\,dex. The latter two were included to assess the effect of the assumed fixed values \logg\ and [M/H] on the \teff\ obtained from the SED. The four individual contributions to the uncertainties in \teff\ were added in quadrature, but overall uncertainty is mostly dominated by the formal least-squares error.

%&&&&&&&&&&&&&&&&&&&&&&&&&&&&&&&&&&&&&&&&&&&&&&&&&&&&&&&&&&&&&&&&&&&&&&&&&&&&&&&&&&&&&&&&&&&&&&&&&&&&&&&&&&&&&&&&&&

\item[(iv)] {\it Balmer lines} -- The values of \teff\ were obtained using the sensitivity of Balmer lines to this parameter, following the method proposed by \citet{catanzaro1}.
On the other hand, hydrogen lines are not good indicators of \logg\ for effective temperatures lower than about $8000$\,K. 
Therefore, for almost all stars but two (KIC\,9828226 and KIC\,11044547), the initial surface gravity was assumed to be equal to $4.0$\,dex.
We used an iterative approach to minimize the differences between observed and theoretical H$\delta$, H${\gamma}$ and H$\beta$ profiles \citep[see][]{catanzaro1}.
The effective temperature obtained from the Balmer lines was then revisited when a \logg\ value from the Fe lines was available. 

The uncertainties in $\log g$ and continuum placement were accounted for when estimating the uncertainty in the derived values.
First, we take into account the differences in the determined \teff\ values from separate Balmer lines.
Next, we check the influence of the assumed $\log g$ value on the final $T_{\rm eff}$ value.
For most stars, the obtained uncertainties equal $100$\,K (this is the step of \teff\ in the grid of synthetic fluxes).
The bigger errors are caused by incorrect normalisation (see Table\,\ref{parameters1}).\\

%&&&&&&&&&&&&&&&&&&&&&&&&&&&&&&&&&&&&&&&&&&&&&&&&&&&&&&&&&&&&&&&&&&&&&&&&&&&&&&&&&&&&&&&&&&&&&&&&&&&&&&&&&&&&&&&&&&

\item[(v)] {\it Iron lines} -- The atmospheric parameters, chemical compositions, and \vsini\ were determined through the analysis of lines of neutral and ionised iron.
In general, we required that the abundances obtained from \ion{Fe}{i} and \ion{Fe}{ii} lines yield the same result.
The detailed analysis of the iron lines was fully discussed in Paper\,I. In brief, we proceeded according to the following scheme:\\
-- microturbulence \micro\ was adjusted until there was no correlation between iron abundances and line depths for the \ion{Fe}{i} lines;\\
-- \teff\ was changed until there was no trend in the abundance versus excitation potential for the \ion{Fe}{i} lines; and\\
-- \logg\ was obtained by requiring the same \ion{Fe} abundance from the lines of both \ion{Fe}{ii} and \ion{Fe}{i} ions.\\
For hotter stars, with more \ion{Fe}{ii} than \ion{Fe}{i} lines visible in the spectra, the \ion{Fe}{ii} lines were used to determine \teff\ and \micro. 
This approach is limited by stellar rotation, since line blending grows rapidly with \vsini. 
In cases where the analysis of individual lines was impossible we assumed that the iron abundances obtained from different spectral regions
are the same for the correct \teff, \logg\ and \micro\ values. 
The analysis relies on the same physics, regardless of rotation (see Paper\,I for further explanation). 

We analysed the metal lines using spectrum synthesis, relying on an efficient least-squares optimisation algorithm (see Paper\,I). 
The chosen method allows for a simultaneous determination of various parameters influencing stellar spectra. The synthetic spectrum depends on \teff, \logg, \micro, \vsini, 
and the relative abundances of the elements. Due to the fact that all atmospheric parameters are correlated,
the \teff, \logg\ and \micro\ parameters were obtained prior to the determination of chemical abundances.
The remaining parameters (chemical abundances and \vsini) were determined simultaneously.
The \vsini\ values were determined by comparing the shapes of observed metal line profiles with the computed profiles, as shown by \citet{gray}.
Chemical abundances and \vsini\ values were determined from many different lines or regions of the spectrum.
In the final step of the analysis, we derived the average values of \vsini\ and abundances of all the chemical elements considered for a given star.

We used atmospheric models (plane-parallel, hydrostatic and radiative equilibrium) and synthetic spectra computed with the line-blanketed, local thermodynamical 
equilibrium (LTE) code {\small\sc ATLAS9} \citep{1993KurCD..13.....K}.
The grid of models was calculated for effective temperatures between $5000$ and $12000$\,K with a step of $100$\,K, 
surface gravities from $2.0$ to $4.6$\,dex with a step of $0.1$\,dex, microturbulence velocities between $0.0$ and $6.0$\,km\,s$^{-1}$ with a step of $0.1$\,km\,s$^{-1}$, 
and for metallicities of $0.0$, $-0.5$ and $-1.0$. The synthetic spectra were computed with the {\small\sc SYNTHE} code \citep{1993KurCD..18.....K}. 
Both codes, {\small\sc ATLAS9} and {\small\sc SYNTHE}, were ported to GNU/Linux by \citet{sbordone} and are available online\footnote{http://wwwuser.oats.inaf.it/castelli/}. 
We used the line list of \citet{castelli}. 
In our method we take into account all elements that show lines in the chosen spectral region.
Elements with little or no influence are assumed to have solar abundances or abundances typical for the analysed star. 
The stellar rotation and signal-to-noise ratio of the spectrum dictate which elements will be analysed.
The atmospheric parameters obtained from our analysis are given in Table\,\ref{parameters1}, and the abundances of analysed elements are given in Table\,B1.%\ref{abundance-table}.

The derived atmospheric parameters and chemical abundances are influenced by errors from a number of sources:\\
-- several assumptions, e.g., plane parallel geometry (1-D), and hydrostatic equilibrium \citep{1993KurCD..13.....K},
made in the atmosphere models used to compute the synthetic spectra; adopted atomic data \citep[see][]{kurucz-alllines};
non-LTE effects for \ion{Fe}{i} and \ion{Fe}{ii} lines \citep[see][]{2011mast.conf..314M}. These error sources have already been discussed in Paper\,I.\\
-- the quality (resolution, S/N) and wavelength range of the observed spectra, as well as their normalisation can influence the results.
The difficulty of normalisation increases for heavily blended spectra of stars with \vsini\ higher than about $100$\,\kms, even for high-resolution and high S/N data.
This problem was also discussed by \citet{2016MNRAS.456.1221R} and \citet{2016MNRAS.458.2307K}. For a few stars we have FIES and HERMES spectra at our disposal.
The HERMES data were analysed in the Paper\,I. The S/N of the data is similar, but the resolution differs significantly.
The HERMES spectra have a resolving power of about $90\,000$, compared to $24\,000$ (low resolution) or $46\,000$ (medium resolution) for FIES.
Both HERMES and FIES spectra are available for KIC\,2694337, KIC\,9828226 (low resolution FIES spectra) and KIC\,4840675 (SB2), KIC\,5988140,
KIC\,7119530, KIC\,7767565, KIC\,8975515 (SB2), KIC\,9970568 (medium resolution FIES spectra).
For these stars we can compare the results obtained from different sets of spectra.
We found that the differences in \teff\ and \logg\ do not exceed $100$\,K and $0.1$\,dex, respectively.
The biggest differences in \micro\ were obtained for KIC\,2694337 ($0.4$\,\kms), KIC\,7119530 ($0.6$\,\kms), and KIC\,9970568 ($0.3$\,\kms; but in agreement within the error bars).
This is due to the effect of different lines used for analysis.
For the other stars the discrepancies were no more than $0.1$\,\kms.
Both \citet{2016MNRAS.456.1221R} and \citet{2016MNRAS.458.2307K} obtained similar results. 
\end{itemize}

The chemical abundances are affected by uncertainties in atmospheric parameters.
For example, an error of \teff\ equal to $100$\,K leads to changes in abundances smaller than $0.1$\,dex in most cases.
Similarly, an error of \logg\ equal to $0.1$\,dex changes chemical abundances by about $0.05$\,dex, and a $0.1$\,\kms\ uncertainty in \micro\ changes abundances by no more than $0.15$\,dex.
The uncertainty of the microturbulence velocity increases with \vsini, from $0.1$\,\kms\ for small and moderate velocities, to $0.4$\,\kms\ for rapid rotators.
Chemical abundances determined for rapidly rotating stars have the largest uncertainty.
The combined errors of chemical abundances were calculated as in Paper\,I. In most cases these errors are lower than $0.20$\,dex.
The \vsini\ values obtained from HERMES ans FIES spectra are consistent.

%&&&&&&&&&&&&&&&&&&&&&&&&&&&&&&&&&&&&&&&&&&&&&&&&&&&&&&&&&&&&&&&&&&&&&&&&&&&&&&&&&&&&&&&&&&&&&&&&&&&&&&&&&&&&&&&&&&

\begin{landscape}
\begin{table}
\caption{Atmospheric parameters of the investigated \kepler\ A- and F-type stars. Photometric values (H2014 and KIC), those obtained from spectral energy distributions (Na \& SED)
and from analysis of metal and Balmer lines are presented. The $1\sigma$ uncertainties are given for effective temperatures from SED and for 
\vsini\ and $\log\epsilon{\rm (Fe)}$ values from metal lines.}
\label{parameters1}
\begin{scriptsize}
\begin{tabular}{crcccc|cc|crcr@{$\,\pm\,$}lc}
\toprule
KIC          &\multicolumn{4}{c}{\hrulefill \,H2014 \& KIC\,\hrulefill}
             &\multicolumn{2}{c}{\hrulefill \,Na\,D \& SED\,\hrulefill}
             &\multicolumn{1}{c}{\hrulefill Balmer lines\,\hrulefill}
             &\multicolumn{6}{c}{\hrulefill \,Fe lines\,\hrulefill}\\
Number       & \teff & \logg & $E(B-V)$ & $\log\epsilon{\rm (Fe)}$ & $E(B-V)$ & \teff\  & \multicolumn{1}{c}{\teff} &\multicolumn{1}{c}{\teff}& \logg  &  \micro\  & \multicolumn{2}{c}{vsini} & $\log\epsilon{\rm (Fe)}$\\
             & [K]   &       & [mag]    &                          & [mag]    &   [K]   & \multicolumn{1}{c}{[K]}   &\multicolumn{1}{c}{[K]}  &        & [\kms]    & \multicolumn{2}{c}{[\kms]} &  \\
\midrule
 2162283  &$  6947 $&$ 4.07 $&$ 0.05 $&$ 7.60 $&$  0.01 $&$  7030 \pm 150 $&$  7100 \pm 100 $&$  7000 \pm 100 $&$  4.1 \pm 0.1 $&$  3.1 \pm 0.1 $&$  42 $&$  2 $&$  7.70 \pm 0.14 $\\
 2694337  &$  7526 $&$ 3.96 $&$ 0.08 $&$ 7.42 $&$  0.06 $&$  7640 \pm 190 $&$  7300 \pm 100 $&$  7400 \pm 100 $&$  4.1 \pm 0.1 $&$  3.0 \pm 0.1 $&$  96 $&$  3 $&$  7.60 \pm 0.07 $\\
 3219256  &$  7493 $&$ 3.59 $&$ 0.06 $&$ 7.48 $&$  0.03 $&$  7550 \pm 180 $&$  7500 \pm 100 $&$  7300 \pm 100 $&$  3.8 \pm 0.1 $&$  3.0 \pm 0.1 $&$  99 $&$  3 $&$  7.40 \pm 0.10 $\\
 3331147  &$  7256 $&$ 3.60 $&$ 0.10 $&$ 7.28 $&$  0.03 $&$  7040 \pm 150 $&$  7000 \pm 100 $&$  7000 \pm 100 $&$  3.7 \pm 0.1 $&$  2.0 \pm 0.1 $&$  63 $&$  2 $&$  7.46 \pm 0.13 $\\
 3429637  &$  7211 $&$ 3.99 $&$ 0.12 $&$ 7.30 $&$  0.02 $&$  7260 \pm 160 $&$  7200 \pm 100 $&$  7300 \pm 100 $&$  3.4 \pm 0.1 $&$  2.9 \pm 0.1 $&$  51 $&$  1 $&$  7.55 \pm 0.08 $\\
 3453494  &$  7766 $&$ 3.64 $&$ 0.09 $&$ 7.38 $&$  0.03 $&$  7830 \pm 200 $&$  7800 \pm 100 $&$  8000 \pm 200 $&$  3.8 \pm 0.2 $&$  2.0 \pm 0.4 $&$ 230 $&$  9 $&$  7.42 \pm 0.11 $\\
 3868420  &$  6782 $&$ 4.08 $&$ 0.06 $&$ 7.24 $&$  0.00 $&$  6090 \pm 110 $&$  7600 \pm 200 $&$  7800 \pm 100 $&$  3.9 \pm 0.1 $&$  2.9 \pm 0.1 $&$  11 $&$  1 $&$  7.93 \pm 0.13 $\\
 4647763  &$  6842 $&$ 3.52 $&$ 0.08 $&$ 7.44 $&$  0.05 $&$  7220 \pm 170 $&$  7100 \pm 100 $&$  7100 \pm 100 $&$  3.8 \pm 0.1 $&$  3.9 \pm 0.1 $&$  69 $&$  2 $&$  7.39 \pm 0.08 $\\
 5880360  &$  7930 $&$ 3.68 $&$ 0.08 $&$ 7.34 $&$  0.05 $&$  8070 \pm 220 $&$  7900 \pm 100 $&$  7700 \pm 200 $&$  3.6 \pm 0.2 $&$  3.5 \pm 0.3 $&$ 187 $&$ 10 $&$  7.18 \pm 0.13 $\\
 5988140  &$  7404 $&$ 3.28 $&$ 0.09 $&$ 7.57 $&$  0.04 $&$  7630 \pm 180 $&$  7800 \pm 200 $&$  7800 \pm 100 $&$  3.8 \pm 0.1 $&$  1.9 \pm 0.1 $&$  52 $&$  2 $&$  7.48 \pm 0.11 $\\
 6123324  &$  6691 $&$ 3.09 $&$ 0.07 $&$ 7.57 $&$  0.05 $&$  7110 \pm 160 $&$  7000 \pm 100 $&$  6800 \pm 200 $&$  3.5 \pm 0.2 $&$  3.9 \pm 0.3 $&$ 189 $&$ 18 $&$  6.63 \pm 0.18 $\\
 6279848  &$  6843 $&$ 3.75 $&$    - $&$ 7.64 $&$  0.03 $&$  7000 \pm 150 $&$  7000 \pm 200 $&$  7100 \pm 100 $&$  4.0 \pm 0.1 $&$  3.2 \pm 0.1 $&$  39 $&$  1 $&$  7.27 \pm 0.09 $\\
 6370489  &$  6302 $&$ 3.92 $&$ 0.02 $&$ 7.30 $&$  0.01 $&$  6300 \pm 120 $&$  6200 \pm 100 $&$  6300 \pm 100 $&$  4.0 \pm 0.1 $&$  1.4 \pm 0.1 $&$   7 $&$  1 $&$  7.12 \pm 0.12 $\\
 6443122  &$  7696 $&$ 3.54 $&$ 0.17 $&$ 6.92 $&$  0.04 $&$  7420 \pm 190 $&$  7100 \pm 100 $&$  7200 \pm 100 $&$  3.8 \pm 0.1 $&$  3.0 \pm 0.1 $&$  83 $&$  3 $&$  7.50 \pm 0.08 $\\
 6670742  &$  7694 $&$ 3.63 $&$ 0.08 $&$ 7.42 $&$  0.02 $&$  7540 \pm 190 $&$  7400 \pm 100 $&$  7300 \pm 200 $&$  3.6 \pm 0.2 $&$  3.2 \pm 0.4 $&$ 276 $&$  8 $&$  7.31 \pm 0.11 $\\
 7119530  &$  7521 $&$ 3.26 $&$ 0.09 $&$ 7.57 $&$  0.07 $&$  8240 \pm 230 $&$  7800 \pm 100 $&$  8200 \pm 200 $&$  3.8 \pm 0.2 $&$  2.5 \pm 0.4 $&$ 240 $&$ 11 $&$  7.35 \pm 0.21 $\\
 7122746  &$  7710 $&$ 4.00 $&$ 0.15 $&$ 7.56 $&$  0.08 $&$  8390 \pm 280 $&$  8200 \pm 100 $&$  8100 \pm 100 $&$  4.1 \pm 0.2 $&$  2.9 \pm 0.2 $&$ 110 $&$  5 $&$  7.46 \pm 0.12 $\\
 7668791  &$  8394 $&$ 3.78 $&$ 0.08 $&$ 7.28 $&$  0.02 $&$  8110 \pm 220 $&$  8200 \pm 100 $&$  8200 \pm 100 $&$  3.8 \pm 0.1 $&$  2.7 \pm 0.1 $&$  51 $&$  2 $&$  7.46 \pm 0.10 $\\
 7748238  &$  7244 $&$ 4.06 $&$ 0.10 $&$ 7.32 $&$  0.05 $&$  7440 \pm 170 $&$  7200 \pm 100 $&$  7400 \pm 100 $&$  3.9 \pm 0.2 $&$  3.0 \pm 0.2 $&$ 129 $&$  4 $&$  7.48 \pm 0.11 $\\
 7767565  &$  8466 $&$ 3.98 $&$    - $&$ 7.71 $&$  0.06 $&$  7810 \pm 200 $&$  7900 \pm 200 $&$  7800 \pm 100 $&$  3.8 \pm 0.1 $&$  2.7 \pm 0.1 $&$  62 $&$  3 $&$  7.75 \pm 0.13 $\\
 7773133  &$  6977 $&$ 3.53 $&$ 0.13 $&$ 7.30 $&$  0.10 $&$  7070 \pm 170 $&$  7100 \pm 100 $&$  7200 \pm 100 $&$  3.9 \pm 0.1 $&$  2.9 \pm 0.1 $&$  41 $&$  2 $&$  7.64 \pm 0.11 $\\
 7798339  &$  6878 $&$ 3.91 $&$ 0.05 $&$ 7.44 $&$  0.01 $&$  7000 \pm 150 $&$  6800 \pm 100 $&$  7000 \pm 100 $&$  3.9 \pm 0.1 $&$  2.5 \pm 0.1 $&$  13 $&$  1 $&$  7.13 \pm 0.12 $\\
 8211500  &$  7619 $&$ 3.93 $&$ 0.04 $&$ 7.60 $&$  0.01 $&$  7730 \pm 180 $&$  7900 \pm 100 $&$  8000 \pm 100 $&$  3.9 \pm 0.1 $&$  2.4 \pm 0.1 $&$  90 $&$  7 $&$  7.56 \pm 0.09 $\\
 8222685  &$  6861 $&$ 4.17 $&$    - $&$ 7.60 $&$  0.01 $&$  7150 \pm 160 $&$  7100 \pm 100 $&$  7100 \pm 100 $&$  3.8 \pm 0.1 $&$  3.2 \pm 0.1 $&$  82 $&$  2 $&$  7.20 \pm 0.07 $\\
 8694723  &$  6120 $&$ 4.10 $&$ 0.03 $&$ 6.91 $&$  0.02 $&$  6370 \pm 120 $&$  6100 \pm 100 $&$  6200 \pm 100 $&$  4.1 \pm 0.1 $&$  1.3 \pm 0.1 $&$   7 $&$  1 $&$  7.02 \pm 0.10 $\\
 8827821  &$  7610 $&$ 3.71 $&$ 0.14 $&$ 7.50 $&$  0.10 $&$  7680 \pm 190 $&$  7400 \pm 100 $&$  7700 \pm 100 $&$  3.7 \pm 0.1 $&$  3.4 \pm 0.1 $&$  87 $&$  2 $&$  7.60 \pm 0.07 $\\
 8881697  &$  7932 $&$ 3.93 $&$ 0.11 $&$ 7.40 $&$  0.05 $&$  7800 \pm 200 $&$  7700 \pm 100 $&$  7900 \pm 200 $&$  3.8 \pm 0.2 $&$  3.3 \pm 0.3 $&$ 184 $&$  4 $&$  7.51 \pm 0.11 $\\
 9229318  &$  7358 $&$ 3.63 $&$ 0.09 $&$ 7.38 $&$  0.09 $&$  7550 \pm 180 $&$  7200 \pm 100 $&$  7400 \pm 100 $&$  4.0 \pm 0.1 $&$  3.0 \pm 0.1 $&$  66 $&$  2 $&$  7.74 \pm 0.08 $\\
 9349245  &$  7914 $&$ 3.72 $&$ 0.05 $&$ 7.06 $&$  0.02 $&$  7880 \pm 210 $&$  8000 \pm 200 $&$  8200 \pm 100 $&$  3.7 \pm 0.1 $&$  2.9 \pm 0.1 $&$  82 $&$  2 $&$  7.92 \pm 0.11 $\\
 9408694  &$  6813 $&$ 3.78 $&$ 0.16 $&$ 7.42 $&$  0.00 $&$  7120 \pm 170 $&$  7200 \pm 100 $&$  7300 \pm 100 $&$  4.0 \pm 0.1 $&$  2.0 \pm 0.1 $&$  16 $&$  1 $&$  7.53 \pm 0.13 $\\
 9410862  &$  6230 $&$ 4.32 $&$ 0.04 $&$ 7.30 $&$  0.04 $&$  6240 \pm 130 $&$  6000 \pm 100 $&$  6100 \pm 100 $&$  3.9 \pm 0.1 $&$  1.3 \pm 0.1 $&$   4 $&$  1 $&$  7.18 \pm 0.13 $\\
 9650390  &$  8572 $&$ 3.77 $&$ 0.10 $&$ 7.30 $&$  0.05 $&$  8510 \pm 290 $&$  8100 \pm 200 $&$  8900 \pm 200 $&$  3.9 \pm 0.2 $&$  3.2 \pm 0.4 $&$ 267 $&$  7 $&$  7.63 \pm 0.11 $\\
 9656348  &$  7177 $&$ 3.90 $&$ 0.14 $&$ 7.57 $&$  0.05 $&$  7600 \pm 180 $&$  7400 \pm 200 $&$  7600 \pm 100 $&$  3.9 \pm 0.1 $&$  3.0 \pm 0.1 $&$  32 $&$  2 $&$  6.83 \pm 0.11 $\\
 9828226  &$  9155 $&$ 4.03 $&$ 0.10 $&$ 7.57 $&$  0.04 $&$  8950 \pm 370 $&$  9000 \pm 200 $&$  9100 \pm 100 $&$  4.1 \pm 0.1 $&$  2.0 \pm 0.1 $&$  99 $&$ 18 $&$  6.39 \pm 0.21 $\\
 9845907  &$  8148 $&$ 4.01 $&$ 0.14 $&$ 7.57 $&$  0.00 $&$  7540 \pm 220 $&$  7900 \pm 200 $&$  7900 \pm 100 $&$  3.9 \pm 0.1 $&$  2.0 \pm 0.1 $&$  13 $&$  1 $&$  7.10 \pm 0.11 $\\
 9941662  &$  9107 $&$ 3.87 $&$ 0.10 $&$ 7.57 $&$  0.03 $&$  6940 \pm 140 $&$  8100 \pm 100 $&$  8200 \pm 100 $&$  4.1 \pm 0.1 $&$  2.7 \pm 0.1 $&$  76 $&$  3 $&$  7.65 \pm 0.06 $\\
 9970568  &$  8035 $&$ 3.71 $&$ 0.11 $&$ 7.40 $&$  0.10 $&$  8080 \pm 220 $&$  7800 \pm 100 $&$  7800 \pm 200 $&$  4.0 \pm 0.2 $&$  2.1 \pm 0.4 $&$ 250 $&$ 11 $&$  7.35 \pm 0.09 $\\
10030943  &$  6883 $&$ 4.23 $&$ 0.07 $&$ 7.64 $&$  0.19 $&$  7410 \pm 210 $&$  6900 \pm 100 $&$  6900 \pm 100 $&$  4.0 \pm 0.1 $&$  3.8 \pm 0.1 $&$  79 $&$  4 $&$  7.07 \pm 0.10 $\\
10341072  &$  6543 $&$ 4.17 $&$ 0.03 $&$ 7.38 $&$  0.02 $&$  6610 \pm 130 $&$  6700 \pm 300 $&$  6700 \pm 100 $&$  3.8 \pm 0.2 $&$  3.0 \pm 0.2 $&$ 104 $&$  2 $&$  7.65 \pm 0.08 $\\
10661783  &$  8117 $&$ 3.72 $&$ 0.10 $&$ 7.34 $&$  0.01 $&$  7960 \pm 220 $&$  7800 \pm 100 $&$  7900 \pm 100 $&$  4.1 \pm 0.1 $&$  1.8 \pm 0.1 $&$  83 $&$  2 $&$  7.51 \pm 0.12 $\\
10717871  &$  7485 $&$ 3.51 $&$ 0.12 $&$ 7.57 $&$  0.03 $&$  7200 \pm 160 $&$  7200 \pm 100 $&$  7100 \pm 100 $&$  3.9 \pm 0.2 $&$  2.0 \pm 0.2 $&$ 128 $&$  5 $&$  7.45 \pm 0.11 $\\
11044547  &$  9143 $&$ 3.89 $&$ 0.11 $&$ 7.57 $&$  0.08 $&$  9360 \pm 470 $&$  9000 \pm 200 $&$  9200 \pm 100 $&$  4.0 \pm 0.1 $&$  1.2 \pm 0.1 $&$  68 $&$  4 $&$  7.60 \pm 0.13 $\\
11622328  &$  7725 $&$ 4.19 $&$    - $&$ 7.57 $&$  0.06 $&$  8250 \pm 240 $&$  8300 \pm 300 $&$  8300 \pm 100 $&$  3.8 \pm 0.2 $&$  2.8 \pm 0.2 $&$ 144 $&$  5 $&$  7.41 \pm 0.09 $\\
11973705  &$ 11086 $&$ 3.98 $&$ 0.05 $&$ 7.57 $&$  0.01 $&$  7520 \pm 180 $&$  7400 \pm 100 $&$  7500 \pm 100 $&$  3.9 \pm 0.2 $&$  3.4 \pm 0.2 $&$ 114 $&$  7 $&$  6.52 \pm 0.14 $\\
\bottomrule
\end{tabular}

Results using TD-1: KIC\,3219256: $7750\pm290$\,K; KIC\,3429637: $7340\pm330$\,K; KIC\,7119530: $8380\pm220$\,K; KIC\,7668791: $8580\pm240$\,K;
KIC\,7798339: $7110\pm290$\,K; KIC\,8211500: $8040\pm290$\,K; KIC\,9349245: $8030\pm200$\,K; KIC\,9650390: $8870\pm240$\,K.\\

Comments on individual stars:\\
KIC\,3868420 (WDS19434+3855): 10.35+10.61\,mag pair, separation 2.5" plus wide (26") 11.71\,mag companion. 2MASS photometry is unreliable. Some photometry is for
the individual stars (e.g. Tycho) but other is for the combined light of the close pair. Insufficient photometry of the individual stars to obtain SED fits
for each star. HADS pulsator, so photometry will sample star at different phases and brightnesses.\\
KIC\,7767565 (WDS19457+4330): close pair 9.50+10.0\,mag, separation 0.3". Results for combined light.\\
KIC\,8211500 (WDS18462+4408): 8.19+11.28\,mag, separation 0.7". Results for combined light, but large magnitude difference means that the faint companion
should not affect the results. \\
KIC\,9408694 (V2367\,Cyg): HADS pulsator, so photometry will sample star at different phases and brightnesses. \\
KIC\,9941662 (Kepler-13, WDS19079+4652): 10.35+10.48\,mag, pair separation 1.1". Photometry is for combined light of the pair. Literature shows that both stars
are similar magnitudes to within $0.1\sim0.2$\,mag. SED fit may represent the average of the two stars, but not the individual stars. Fits are therefore not
considered reliable.

\end{scriptsize}
\end{table}
\end{landscape}

%&&&&&&&&&&&&&&&&&&&&&&&&&&&&&&&&&&&&&&&&&&&&&&&&&&&&&&&&&&&&&&&&&&&&&&&&&&&&&&&&&&&&&&&&&&&&&&&&&&&&&&&&&&&&&&&&&&

\begin{figure}
\centering
\includegraphics[width=8cm,angle=0]{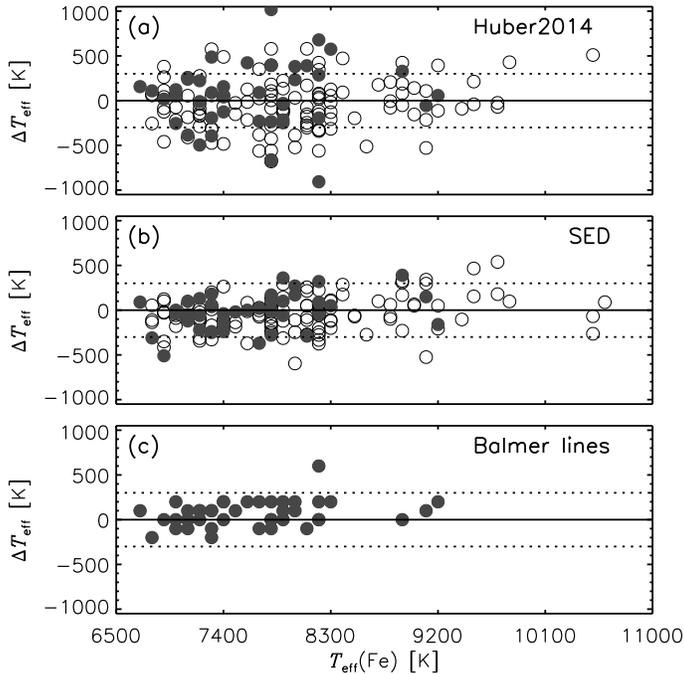}
\caption[]{Comparison of effective temperatures derived by different methods. Filled circles represent the results obtained in this work, open circles are used for Paper\,I stars. In panel (a) the differences $\Delta$\teff\,$=$\teff({\sc Fe})$-$\teff({\sc H2014}) are shown. In panel (b) $\Delta$\teff\,$=$\teff({\sc Fe})$-$\teff({\sc SED}) are plotted. In panel (c) $\Delta$\teff\,$=$\teff({\sc Fe})$-$\teff({\sc Balmer}) are shown.}
\label{tempcomparison}
\end{figure}

%&&&&&&&&&&&&&&&&&&&&&&&&&&&&&&&&&&&&&&&&&&&&&&&&&&&&&&&&&&&&&&&&&&&&&&&&&&&&&&&&&&&&&&&&&&&&&&&&&&&&&&&&&&&&&&&&&&

\begin{figure}
\centering
\includegraphics[width=12cm,angle=0]{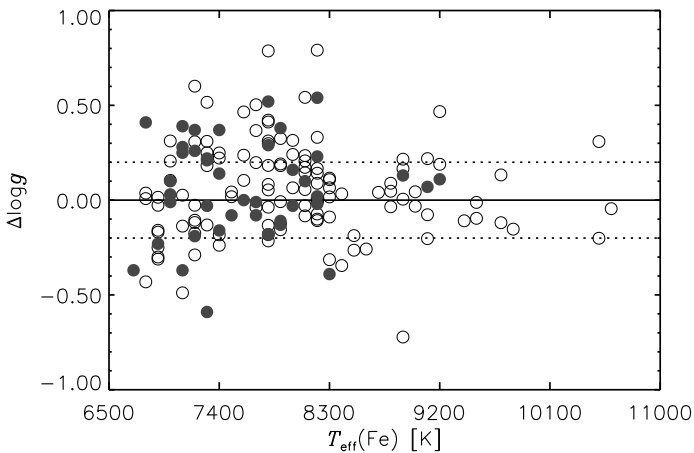}
\caption[]{Differences $\Delta \log g = \log g{\rm (Fe)} - \log g{\rm(H2014)}$ as a function of \teff({\sc Fe}). The stars indicated by open circles are taken from Paper\,I. The filled circles show the results obtained in this work. The dotted horizontal lines indicate a difference of $\pm$ 0.2\,dex.}
\label{loggcomparison}
\end{figure}

%&&&&&&&&&&&&&&&&&&&&&&&&&&&&&&&&&&&&&&&&&&&&&&&&&&&&&&&&&&&&&&&&&&&&&&&&&&&&&&&&&&&&&&&&&&&&&&&&&&&&&&&&&&&&&&&&&&

\section{Discussion of the results}
\label{sec:results}

\subsection{Atmospheric parameters}

Here we discuss the consistency of our atmospheric parameters from the FIES spectra with those from two other sources: the H2014 catalogue, and SEDs.
The $v\sin i$ values are compared with typical values for A and F stars. All parameters are presented in the Table\,\ref{parameters1}. \\

\noindent{\it Effective temperature}\\
The effective temperatures given by H2014 and determined using the methods described in Sect.\,\ref{sec:methods} are presented in Table\,\ref{parameters1} and compared in Fig.\,\ref{tempcomparison}.
In this figure, the differences between \teff({\sc Fe}) obtained from the analysis of the high-resolution spectra and values taken from H2014, \teff({\sc H2014}), from SED fitting, \teff(SED),
and from  the Balmer lines analysis, \teff({\sc Balmer}), are shown as a function of \teff({\sc Fe}). The biggest differences are for \teff({\sc H2014}) values.
The H2014 catalogue is a compilation of literature values for atmospheric properties (\teff, {\logg} and [Fe/H]) derived from different methods, including photometry,
spectroscopy, asteroseismology, and exoplanet transits. For the stars considered here, the H2014 values have been determined with large uncertainties via photometry.
For most of the stars, the values of \teff({\sc Fe})$-$\teff({\sc H2014}) agree to within $\pm300$\,K.
The biggest differences are obtained for chemically peculiar stars KIC\,3868420 (Am star from high resolution spectra analysis;HADS pulsator),
KIC\,7767565 (Am), KIC\,9941662 (Am:), KIC\,11973705 ($\lambda$\,Boo), and for rapidly rotating stars with \vsini\ exceeding $100$\,\kms, e.g. KIC\,11622328 and KIC\,7119530. 

Figure\,\ref{tempcomparison}\,(b) compares \teff({\sc Fe}) with \teff(SED). The consistency between the results of these two methods is good.
To compare the effective temperatures obtained with different methods we use the Kendall's rank correlation coefficient that measures the strength of dependence between two variables.
There is no significant trend of the differences [\teff({\sc Fe})$-$\teff(SED)] with \teff({\sc Fe}). The Kendall's rank correlation coefficient is $0.16$.
The differences exceed $500$\,K only for KIC\,10030943 and KIC\,9941662. KIC\,10030943 is an eclipsing binary star.
The difference \teff({\sc Fe})$-$\teff(SED) may be the result of the influence of the secondary star on the photometric indices and/or on the spectrum.
KIC\,9941662 (\textit{Kepler}-13) is also a binary system, with a magnitude difference $\Delta m=0.13$ and 1.1" separation.
The photometry used to construct the SED is the combined light of the pair.
The \teff({\sc SED}) analysis supports the claimed accuracy of our spectroscopic temperatures because the stars are similar.

The differences \teff({\sc Fe})$-$\teff(Balmer) are in all cases smaller than 300\,K (see Fig.\,\ref{tempcomparison}\,c).\\

\noindent{\it Surface gravities}\\
Figure\,\ref{loggcomparison} compares the \logg({\sc Fe}) values obtained from high-resolution HERMES (open circles, Paper\,I) and FIES (filled circles, this work)
spectroscopy with those taken from the H2014 catalogue, \logg(H2014). For most stars the surface gravities are consistent to within $\pm0.2$\,dex.
There is no correlation between the differences [\logg({\sc Fe})$-$\logg(H2014)] and effective temperature or surface gravity.
The values in the H2014 catalogue were mostly calculated from photometry. The high-resolution spectra are much more sensitive to \logg\ than the photometric indices.
Therefore, the spectroscopic surface gravities determined from iron lines should be adopted for the subsequent analysis of these stars.\\

%&&&&&&&&&&&&&&&&&&&&&&&&&&&&&&&&&&&&&&&&&&&&&&&&&&&&&&&&&&&&&&&&&&&&&&&&&&&&&&&&&&&&&&&&&&&&&&&&&&&&&&&&&&&&&&&&&&

\begin{figure}
\centering
\includegraphics[width=8cm,angle=0]{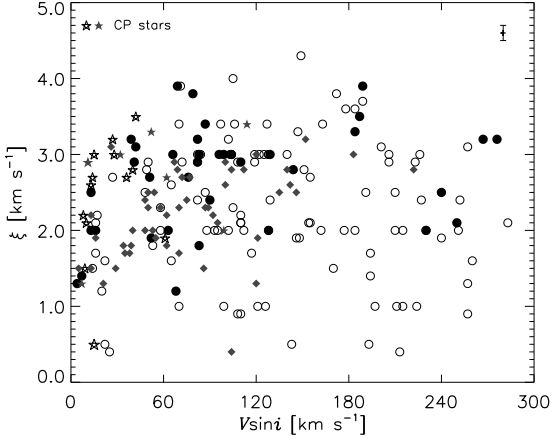}
\caption[]{Microturbulent velocities as a function of the projected rotational velocity.
The CP stars from Paper\,I are shown as open stars, whereas the non-CP stars are indicated as open circles.
The CP and non-CP objects from this work are presented as filled stars and circles, respectively.
Diamonds show results from \citet{2016MNRAS.458.2307K}.
Although microturbulence shows no dependence on \vsini, it is determined to higher precision in slow rotators.}
\label{micro-calc-vsini}
\end{figure}

%&&&&&&&&&&&&&&&&&&&&&&&&&&&&&&&&&&&&&&&&&&&&&&&&&&&&&&&&&&&&&&&&&&&&&&&&&&&&&&&&&&&&&&&&&&&&&&&&&&&&&&&&&&&&&&&&&&

\begin{figure}
\centering
\includegraphics[width=8cm,angle=0]{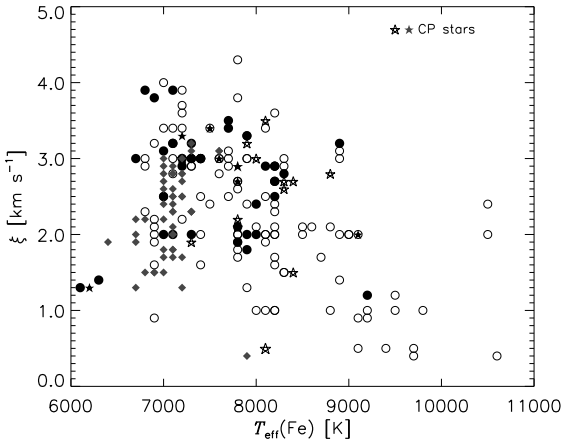}
\caption[]{Microturbulent velocities as a function of effective temperature. 
The CP stars from Paper\,I are shown as open stars, whereas the non-CP stars are indicated as open circles.
The CP and non-CP objects from this work are presented as filled stars and circles, respectively. Diamonds show results from \citet{2016MNRAS.458.2307K}.}
\label{micro-calc}
\end{figure}

%&&&&&&&&&&&&&&&&&&&&&&&&&&&&&&&&&&&&&&&&&&&&&&&&&&&&&&&&&&&&&&&&&&&&&&&&&&&&&&&&&&&&&&&&&&&&&&&&&&&&&&&&&&&&&&&&&&

\noindent{\it Microturbulent velocity}\\
One of the important key parameters in the abundance analysis is microturbulence \micro. 
Microturbulent velocity originates from non-thermal velocities in the stellar atmosphere at scales shorter than the mean-free-path of a photon. 
It has a significant effect on the spectral lines of A and F stars, therefore its determination is important in abundance analyses, especially when using strong line features.

The microturbulent velocities obtained in our analyses lie within the range from about $1.2$ to $3.9$\,\kms. 
Microturbulent velocities are calculated using Fe lines of various strengths, leading to a dependence on spectral quality. 
The best data are characterized by a high signal-to-noise and high resolution. Low S/N or/and resolution prevents weak lines from being analysed. 
Additionally, the precision of the calculated \micro\ values depends strongly on \vsini.
At high $v\sin i$, line blending obscures weak lines. The uncertainties of \micro\ listed in Table\,\ref{parameters1} increase with the increasing \vsini\ value.

To enlarge the sample, the results from Paper\,I and \citet{2016MNRAS.458.2307K} were taken into account.
\citet{2015MNRAS.450.2764N} analysed stars with a broad range of \vsini\ values and found \micro\ 
within the range 2--4\,\kms\ for stars with $7000 <$ \teff\ $\leq 8000$\,K, decreasing to $2$\,\kms\ at $8000 <$ \teff\ $\leq 9000$\,K,
and decreasing further to $0.5$--$1$\,\kms\ for \teff\ $> 9000$\,K.
With a larger sample of stars analysed in a consistent way, we re-examine the dependence of \micro\ on \teff, \logg,
and the influence of \vsini\ and chemical composition on the determined values.

The absence of a correlation between \micro\ and \vsini\ (Fig.\,\ref{micro-calc-vsini}) shows that microturbulence is still well determined across a wide range of \vsini\,
despite the aforementioned challenges, and that there is no physical connection between the quantities, unless that connection is anti-correlated with a systematic misestimation.
There is also no correlation of \micro\ and iron abundance, however the analysed objects fall within a narrow range of iron abundances,
in which most stars have $\log \epsilon({\rm Fe})$ between about $7.0$ and $8.0$.

In Fig.\,\ref{micro-calc} the microturbulences are shown as a function of effective temperature.
Here, a broad maximum in \micro\ is observed around $7000$ -- $8500$\,K.
These results are consistent with the previous determinations of microturbulence for A and F stars.
\citet{1970A&A.....4..291C} first found that microturbulence varies with effective temperature, from $2$\,\kms\ for early A stars up to $4$\,\kms\ for late A stars,
and $2$\,\kms\ for mid F stars. 
Since then the dependence of \micro\ with \teff\ has been investigated in many studies
\citep[][Paper\,I]{1992A&AS...95...41C, 1993A&A...275..101E, 2001AJ....121.2159G, 2004IAUS..224..131S, takeda2008, 2014psce.conf..193G}.
The large dispersion in the results was found in all mentioned works except for \citet{takeda2008}.

The relation between \micro\ and \logg\ was also checked. This dependence was analysed by \citet{2001AJ....121.2159G} for the broad range of \logg\ from $1.0$ to $5.0$.
They discovered the clear correlations between \logg\ and \micro\ for spectral types from A5 to G2.
The microturbulent velocity in giants and supergiants is generally larger than in dwarfs.
Our targets occupy a narrow \logg\ range on or near the main-sequence, preventing verification of the \citet{2001AJ....121.2159G} result. 

As in Paper\,I, the determined microturbulent velocities of Am stars are similar to those of non-CP stars. 
The distribution of microturbulent velocities among the Am stars is indistinguishable from that of normal stars.
This is in stark contrast to results presented by \citet{landstreet2009},
who claimed that Am stars have higher microturbulent velocities, amid criticism of inadequate sampling \citep{murphy2014}.

%&&&&&&&&&&&&&&&&&&&&&&&&&&&&&&&&&&&&&&&&&&&&&&&&&&&&&&&&&&&&&&&&&&&&&&&&&&&&&&&&&&&&&&&&&&&&&&&&&&&&&&&&&&&&&&&&&&

\begin{figure*}
\centering
\includegraphics[width=18cm,angle=0]{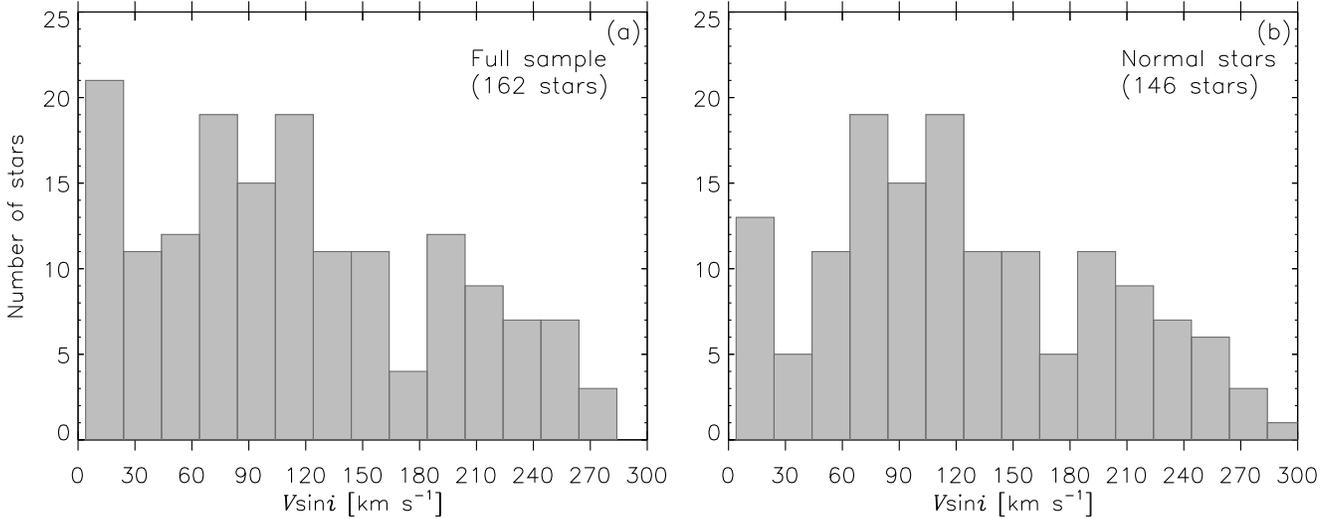}
\caption[]{The distribution of rotational velocities for all stars (panel {\it a}) and for all non-CP stars (panel {\it b}) analysed here and in Paper\,1.}
\label{vsini2}
\end{figure*}

%&&&&&&&&&&&&&&&&&&&&&&&&&&&&&&&&&&&&&&&&&&&&&&&&&&&&&&&&&&&&&&&&&&&&&&&&&&&&&&&&&&&&&&&&&&&&&&&&&&&&&&&&&&&&&&&&&&

\subsection{Projected rotation velocity}
\label{sec:vsini}

The stars analysed here have projected rotational velocities ranging from 4 to 270\,\kms.
In Fig.\,\ref{vsini2}\,(a) the distribution of the \vsini\ values obtained here and in Paper\,I is shown.

Stars with \vsini\ lower than 30\,\kms\ are considered here as slowly rotating stars.
There are seven such objects in the sample analysed here, including KIC\,3868420 (A9\,Vs) showing weakness of Ca lines and
KIC\,8694723 (hF6\,mF2\,gF5\,V) which shows metal underabundances.
There are four single non-CP stars with low \vsini: KIC\,7798339 (F1.5\,V), KIC\,9845907 (A4\,IVs), KIC\,9410862 (F6\,V), and KIC\,6370489 (F5.5\,V).
The slow rotation of F stars is also surprising, if the spectral type is earlier than F5, where the Kraft break is seen \citep{1967ApJ...150..551K}.
Early F stars have a continuous rotational velocity distribution with late A stars. One of the stars, KIC\,9408694 (F0\,IVs + ?) is a member of binary system. 

Figure\,\ref{vsini2}\,(b) shows the distribution of \vsini\ for all non-CP stars analysed here and in Paper\,I.
The mean value of \vsini\ for the whole sample is $125$\,\kms, and the median is $110$\,\kms.
These results are consistent with the projected rotational velocities of A stars available in the literature.
\citet{abt2000} discovered that most of the non chemically peculiar A0--F0 main-sequence stars have \vsini\ values greater than $120$\,\kms.
The average projected rotation velocity of about $130$\,\kms\ was determined by e.g. \citet{royer2009} and \citet{abt2009}.
Our results are consistent with their conclusions. 

The occurrence of non-CP A stars with low \vsini\ has been discussed in Paper\,I. Various authors have tried to explain the origin of such stars.
\citet{abt2009} considered binarity and the amount of time necessary for slow rotators to become Ap or Am stars by a diffusion mechanism \citep*[][]{2006ApJ...645..634T, 2013EAS....63..199M}.
A more probable reason for the low observed \vsini\ in some A stars is the inclination effect, as for Vega \citep[][]{gray1985vega, 1994ApJ...429L..81G, abt2009, 2010ApJ...712..250H}.
The spectroscopically obtained \vsini\ is a projection of the equatorial velocity along the line-of-sight,
so stars classified as slow rotators can be in fact fast rotators seen at low inclination angle.
To investigate the effect of low inclination on the observed spectrum of a fast rotator, 
high-resolution and very high signal-to-noise (S/N\,$\sim 1000$) spectra are necessary \citep[e.g.][]{royer2012}. 

%&&&&&&&&&&&&&&&&&&&&&&&&&&&&&&&&&&&&&&&&&&&&&&&&&&&&&&&&&&&&&&&&&&&&&&&&&&&&&&&&&&&&&&&&&&&&&&&&&&&&&&&&&&&&&&&&&&

\begin{table}
\centering
\caption{Average abundances ($\log\epsilon(\rm El)$), standard deviations and number of stars used for abundance calculations
for non-CP, Am, and $\lambda$\,Boo stars, compared to the reference solar values of \citet{asplund2009}.}
\label{abundance-average-tab}
\begin{tabular}{c||lll||c}
\toprule
El.     &  non-CP                & Am                    & $\lambda$\,Boo        &  Solar  \\
        &  stars                 & stars                 & stars                 &  values \\
\midrule
     C  & $8.37,\,0.25\,(37)$ & $   8.26,\, 0.28\,(4) $& $   8.40,\,0.08\,(3) $& $8.43$\\
     N  & $8.17,\,0.30\,( 7)$ & $   8.72\,        (1) $& $   8.31\,       (1) $& $7.83$\\
     O  & $8.84,\,0.25\,(31)$ & $   8.68,\, 0.32\,(4) $& $   8.69,\,0.20\,(3) $& $8.69$\\
    Na  & $6.45,\,0.43\,(30)$ & $   6.59,\, 0.12\,(4) $& $   6.20\,       (1) $& $6.24$\\
    Mg  & $7.69,\,0.24\,(37)$ & $   7.71,\, 0.27\,(4) $& $   6.95,\,0.51\,(3) $& $7.60$\\
    Al  & $6.38,\,0.36\,( 7)$ & $   6.73,\, 0.08\,(2) $& $                    $& $6.45$\\
    Si  & $7.50,\,0.24\,(37)$ & $   7.71,\, 0.05\,(4) $& $   6.96,\,0.33\,(3) $& $7.51$\\
     S  & $7.38,\,0.27\,(29)$ & $   7.39,\, 0.22\,(4) $& $   6.58\,       (1) $& $7.12$\\
    Ca  & $6.42,\,0.26\,(37)$ & $   6.09,\, 0.47\,(4) $& $   5.85,\,0.37\,(3) $& $6.34$\\
    Sc  & $3.17,\,0.28\,(37)$ & $   2.97,\, 0.29\,(4) $& $   2.58,\,0.59\,(3) $& $3.15$\\
    Ti  & $5.01,\,0.21\,(37)$ & $   5.16,\, 0.17\,(4) $& $   4.11,\,0.26\,(3) $& $4.95$\\
     V  & $4.33,\,0.40\,(28)$ & $   4.80,\, 0.15\,(4) $& $   4.07\,       (1) $& $3.93$\\
    Cr  & $5.62,\,0.22\,(37)$ & $   6.09,\, 0.10\,(4) $& $   4.79,\,0.19\,(3) $& $5.64$\\
    Mn  & $5.36,\,0.29\,(37)$ & $   5.41,\, 0.25\,(4) $& $   4.65\,       (1) $& $5.43$\\
    Fe  & $7.40,\,0.24\,(37)$ & $   7.76,\, 0.13\,(4) $& $   6.57,\,0.24\,(3) $& $7.50$\\
    Co  & $5.43,\,0.64\,(16)$ & $   5.49,\, 0.13\,(2) $& $   4.73,\,      (1) $& $4.99$\\
    Ni  & $6.24,\,0.31\,(37)$ & $   6.68,\, 0.20\,(4) $& $   5.77,\,0.13\,(2) $& $6.22$\\
    Cu  & $4.15,\,0.44\,(23)$ & $   4.63,\, 0.26\,(3) $& $   4.13,\,      (1) $& $4.19$\\
    Zn  & $4.41,\,0.31\,(26)$ & $   4.86,\, 0.31\,(4) $& $   3.98,\,      (1) $& $4.56$\\
    Ga  & $2.94,\,0.69\,( 4)$ & $   3.70,\, 0.45\,(2) $& $                    $& $3.04$\\
    Sr  & $2.99,\,0.85\,(36)$ & $   3.98,\, 0.09\,(4) $& $   2.49,\,0.66\,(2) $& $2.87$\\
     Y  & $2.37,\,0.38\,(33)$ & $   2.99,\, 0.05\,(4) $& $   1.84,\,0.18\,(2) $& $2.21$\\
    Zr  & $3.03,\,0.34\,(29)$ & $   3.09,\, 0.14\,(4) $& $   2.58\,       (1) $& $2.58$\\
    Ba  & $2.78,\,0.53\,(36)$ & $   3.82,\, 0.14\,(4) $& $   1.51,\,0.25\,(3) $& $2.18$\\
    La  & $1.43,\,0.44\,(22)$ & $   2.39,\, 0.19\,(3) $& $   1.16\,       (1) $& $1.10$\\
    Ce  & $1.97,\,0.34\,(15)$ & $   2.76,\, 0.11\,(3) $& $   1.47\,       (1) $& $1.58$\\
    Pr  & $0.73,\,0.39\,( 6)$ & $   1.67 \,       (1) $& $                    $& $0.72$\\
    Nd  & $1.86,\,0.47\,(18)$ & $   2.48,\, 0.11\,(3) $& $   1.75\,       (1) $& $1.42$\\
    Sm  & $1.28,\,0.54\,( 5)$ & $                     $& $                    $& $0.96$\\
    Eu  & $1.28,\,0.95\,( 5)$ & $   2.01 \,       (1) $& $                    $& $0.52$\\
    Gd  & $1.27,\,0.34\,( 4)$ & $                     $& $                    $& $1.07$\\
    Dy  & $1.11,\,0.50\,( 4)$ & $                     $& $                    $& $1.10$\\
    Er  & $1.49,\,0.14\,( 3)$ & $                     $& $                    $& $0.92$\\
\bottomrule
\end{tabular}
\end{table}

%&&&&&&&&&&&&&&&&&&&&&&&&&&&&&&&&&&&&&&&&&&&&&&&&&&&&&&&&&&&&&&&&&&&&&&&&&&&&&&&&&&&&&&&&&&&&&&&&&&&&&&&&&&&&&&&&&&

\section{Chemical abundances}
\label{sec:abundances}
In Table\,\ref{abundance-average-tab} the average abundances for a sample of non-CP and for all stars are compared with the photospheric solar values of \citet{asplund2009}.
The abundances of C, Mg, Si, Ca, Sc, Ti, Cr and Fe were derived for all stars.
For \ion{O} we used lines that are only slightly affected by non-LTE effects \citep[e.g.][]{przybilla2000} (e.g., 5331, 6157, 6158, and 6456\,\AA).
Lines of heavy elements except for Sr, Y, Zr, Ba, and rare-earth elements are very sparse in spectra of A stars.
Their abundances were investigated only for slowly and moderately rotating stars and in most cases only one or two blends were available.

%&&&&&&&&&&&&&&&&&&&&&&&&&&&&&&&&&&&&&&&&&&&&&&&&&&&&&&&&&&&&&&&&&&&&&&&&&&&&&&&&&&&&&&&&&&&&&&&&&&&&&&&&&&&&&&&&&&

\begin{figure*}
\centering
\includegraphics[width=18cm,angle=0]{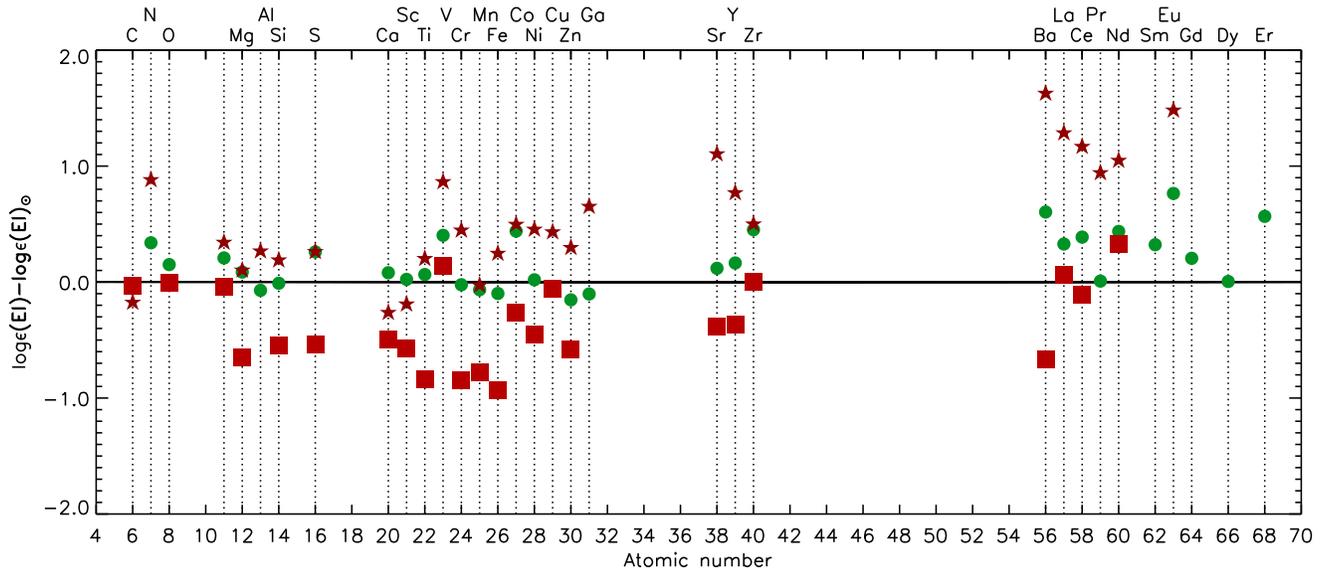}
\caption[]{Average abundances of the non-CP stars (circles), Am (stars) and $\lambda$\,Bo\"{o}tis stars (squares) compared to the solar values of \citet{asplund2009}.}
\label{abundance-average-fig}
\end{figure*}

%&&&&&&&&&&&&&&&&&&&&&&&&&&&&&&&&&&&&&&&&&&&&&&&&&&&&&&&&&&&&&&&&&&&&&&&&&&&&&&&&&&&&&&&&&&&&&&&&&&&&&&&&&&&&&&&&&&

\subsection{Chemically peculiar stars}

In Fig.\,\ref{abundance-average-fig}, the determined average abundances for CP and non-CP stars are compared with the solar abundances of \citet{asplund2009}. 
The average abundances of most of the light and iron-peak elements are similar to the solar values. 
The largest differences were derived for the elements represented by weak blends only.
As expected, the abundances of some elements determined for CP stars are far different from the solar values. 
The CP stars in our sample consist of regular and mild Am stars and $\lambda$\,Bo\"{o}tis candidates.
According to the abundance analysis, five stars were classified as Am:, three single stars KIC\,3868420, KIC\,7767565, KIC\,9941662,
EB system KIC\,2162283 and SB2 star KIC\,3858884, the primary of which is a mild Am star (kA8\,hF0\,mF2.5\,III). 
A detailed abundance analysis of SB2 systems, including KIC\,3858884, will be presented in a forthcoming paper (Catanzaro et al., in preparation),
because disentangling the spectral components from each star demands a different methodology.

KIC\,3868420 was classified initially as A9\,Vs with a very weak \ion{Ca}{ii}\,K line in comparison with the hydrogen line type.
However, no metal-line enhancement was detected under spectral classification.
From the abundance analysis KIC\,3868420 is a typical Am star with underabundant Ca and Sc but enhanced metal lines.
KIC\,7767565 (kA5\,hA7\,mF1\,IV) shows the typical Am abundance pattern except for apparently normal Ca and Sc abundances.
The star was classified as an Am in Paper\,I and the same spectral type was found.
For KIC\,9941662 (kA2\,hA5m\,A7\,(IV)) the spectroscopic data analysis revealed lower Ca and Sc abundances, typical for Am stars.

KIC\,2162283 was classified as a star with a composite Am\,$+$\,F spectrum. The spectral type obtained from hydrogen line is roughly consistent with an A9 giant. 
On the other hand, the extremely strong metal lines match a mid or late-F star.
The star shows overabundances of most iron-peak elements and some heavy elements like Zn, Sr, Zr, and Ba, 
typical for an Am type, but the expected underabundance of Sc is small and Ca has normal abundance, contrary to expectation \citep{gray&corbally2009}.
From spectral classification the spectrum could be interpreted as an Am or Ap star due to the selective enhancements.
The fact that Ca is not weak might appear to rule out an Am type, but the overall metal line morphology is best described by a composite spectrum, 
in which case the \ion{Ca}{II}\,K line would be the mix of a weak Am type and a strong late-F type, and thus appear normal.
Further, the strength of the overall metal line spectrum (in addition to the selective enhancement of some elements) is not typical of Ap stars,
and this star is a known eclipsing binary. 

KIC\,3429637 (hA8\,kF0\,mF0\,IIIas) was previously classified as hA9\,kF2\,mF3 type by \citet{1984ApJ...285..247A} and as F0\,III star by \citet{2011MNRAS.411.1167C}.
KIC\,3429637 is not an Am star. It's a giant with slightly enhanced metal lines.

There are three candidate $\lambda$\,Bo\"{o}tis stars in our sample: KIC\,9656348, KIC\,9828226 and KIC\,11973705.
They are characterised by a weak \specline{Mg}{II}{4481} line and solar abundances of C, N, O, and S \citep{gray&corbally2009}.
The spectrum of KIC\,9656348 (hA9\,kA4\,mA4\,V) has hydrogen lines with deep cores but the core-wing boundaries and the wings themselves are narrow.
Metal lines are also narrow and fit between A3 and A5\,V, hence rotation cannot explain the metal weakness.
KIC\,9828226 (hA2\,kB9.5\,mB9\,Vb) has tremendously broad hydrogen lines in the spectrum, hence the Vb luminosity class.
Its \specline{Mg}{ii}{4481} line is weaker than the metal line type, that is, even weaker than at B9.
Other metals are barely discernible. Hence, KIC\,9828226 is a firm $\lambda$\,Boo candidate.
This star was analysed in Paper\,I, where similar spectral type (hA2\,kA0\,mB9\,Vb) and atmospheric parameters were found.
Its moderate rotational velocity, $99\pm18$\,\kms, allowed us to obtain the abundances of C, O, Mg, Si, Ca, Sc, Ti, Cr, Fe, and Ba.
The chemical abundances were determined on the basis of one or two lines only, except for Ti, Cr and Fe.
All these elements are underabundant. The C and O abundances are solar, typical for $\lambda$\,Bo\"{o}tis stars.
KIC\,11973705 (hF0.5\,kA2.5\,mA2.5\,V) has a typical $\lambda$\,Boo spectrum, with pronounced metal weakness and broad-but-shallow hydrogen line wings.
Its abundances support the $\lambda$\,Boo classification.

In addition, a metal-weak late-type star was found.
KIC\,8694723 was classified as hF6\,mF2\,gF5\,V, with hydrogen lines matching an F6 star, but noticeably weaker metal lines and G-band.

%&&&&&&&&&&&&&&&&&&&&&&&&&&&&&&&&&&&&&&&&&&&&&&&&&&&&&&&&&&&&&&&&&&&&&&&&&&&&&&&&&&&&&&&&&&&&&&&&&&&&&&&&&&&&&&&&&&

\subsection{Correlation of chemical abundances with other parameters}

We used the stars analysed here and those from Paper\,I to search for possible correlations between the abundances of chemical elements and other stellar parameters.
Also in this case we used the Kendall's rank correlation coefficient.
Any correlations or anticorrelations found between abundances and atmospheric parameters are very useful for investigating hydrodynamical processes in stellar atmospheres.
Because of the large sample of stars, we are able to make the reliable analysis of possible correlations between the chemical abundances and \teff, \logg, \micro, \vsini\,
and \ion{Fe} abundances. Only in a few cases was the analysis complicated by large scatter or by having only a small number of stars with measured abundances of a particular element.
This is the case for some of the heavy and rare-earth elements.

Usually, chemically normal stars showed at most mild correlation between element abundances and atmospheric parameters.
Strong positive correlations with \teff\ were only calculated for Zn ($0.43$), and Ce ($0.57$).
In case of Zn the positive correlation can be eliminated by removing the low temperature outliers.
Any other correlations can be explain by small number of abundance calculations for a given element, rendering the correlation determination uncertain.
Simiraly, at mild correlations with \logg\ and \micro\ were found.
As expected, positive correlations with iron abundance were found for almost all elements.
The highest correlation coefficients were obtained for iron-peak elements Ca, Ti, Cr, Ni, Cu, plus Al, La and Ce, all of which have correlation coefficients $>0.30$.
\citet{takeda2008} also found strong positive correlations between the abundances of Si, Ti and Ba, and that of Fe.

For CP stars, mild positive correlations (obtained for Al, S, Mn, Ni, Cu, Zn, Sr, Zr, Ba, La) may again be the result of small number statistics but this time because CP stars are few in number.
\citet{2000ApJ...529..338R} also predicted mild correlations between abundances of some elements (e.g. Ni) and \teff.
Strong correlations were found only Ca ($0.52$) and O, Co, and Ce ($\sim -0.40$).
Strong positive correlations with \logg\ were found for Na, V, Mn, Cu, and Zn ($0.30 - 0.50$).
The strongest positive correlation was found for Ni ($0.62$), while negative were obtained for N ($-0.62$) and Sc ($-0.33$).
The strong positive correlations with \micro\ were found for N ($0.55$), Cr ($0.32$), and Mn ($0.39$),
while negative correlations were derived for C ($-0.38$), and Ba ($-0.41$).
However, values for the smallest \micro\ sometimes differ significantly from the other results. Removing these values erases any existing correlations.
Positive correlations ($>0.30$) with iron abundances were obtained for Na, Si, Ca, Ti, V, Cr, Mn, Ni, and La.
All correlations for CP should be treated with caution, because of the small number of the analysed CP stars.

In the present work, no strong correlations between the abundances of most elements and \vsini\ were found among normal stars.
Stronger correlations were found only for Co ($0.54$), Nd ($0.41$), and Sr ($-0.40$). 
Only for Co and Sr can the correlation be considered as reliable, since for the other elements the number of stars considered is less than $20$.
For chemically peculiar stars strong positive correlations were found for Na, Si, Ti, Mn, and Co, with correlation coefficients higher than $0.40$.
Comparing samples with A and F stars, we found a larger scatter of abundances for hotter stars, especially for stars with \vsini\ higher than 150\,\kms.

The same trend was obtained for the abundances of A and F stars in the Hyades \citep{2010A&A...523A..71G}, Pleiades \citep{2008A&A...483..567G},
and Coma Berenices open clusters \citep{2008A&A...479..189G}. 
The correlations between abundances and \vsini\ were examined by \citet{fossati2008} and \citet{takeda2008}.
\citet{fossati2008} investigated $23$ normal A and F stars and found no correlation between abundances and projected rotational velocity.
A larger sample was studied by \citet{takeda2008}, who discovered negative correlations for C, O, Ca
and strong positive correlations for the Fe-peak elements plus Y and Ba, with rotational velocity.

%&&&&&&&&&&&&&&&&&&&&&&&&&&&&&&&&&&&&&&&&&&&&&&&&&&&&&&&&&&&&&&&&&&&&&&&&&&&&&&&&&&&&&&&&&&&&&&&&&&&&&&&&&&&&&&&&&&

%----------
\begin{figure*}
\centering
\includegraphics[width=16cm,angle=0]{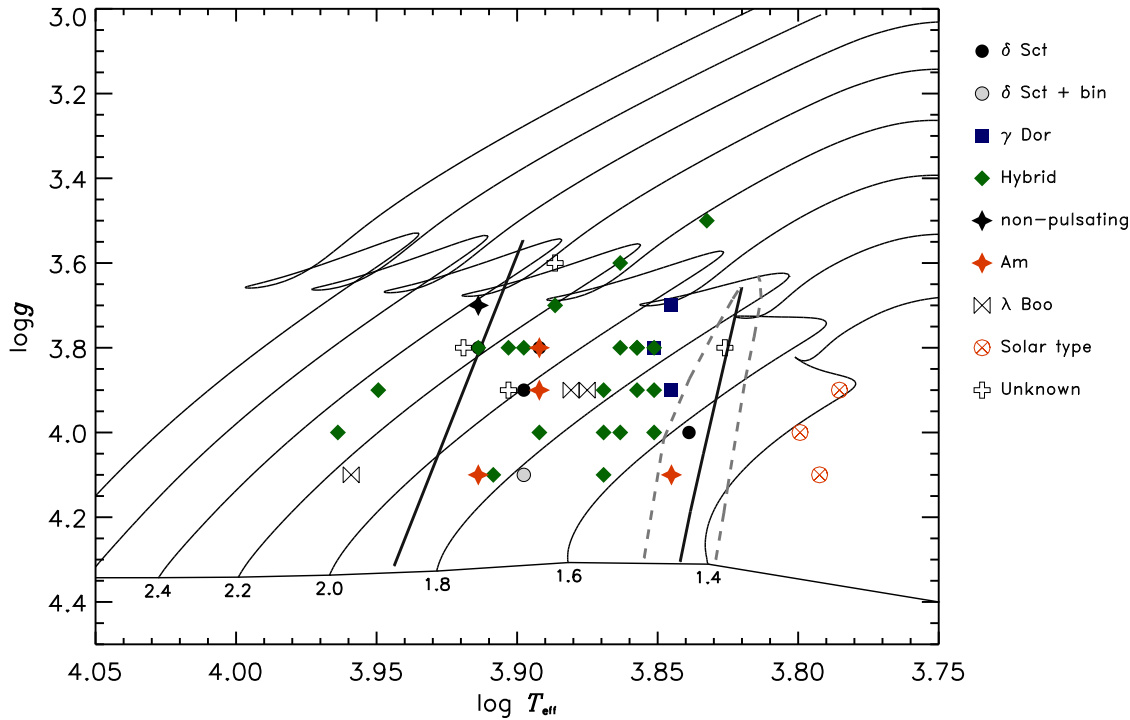}
\caption[]{$\log$\,\teff -- \logg\ diagram for the stars analysed in the current work.
Stars identified as $\delta$\,Sct, $\gamma$\,Dor, hybrids, non-pulsating or CP objects are plotted with different symbols (see legend).}
\label{hrall}
\end{figure*}

%----------

\section{Variability and spectroscopic H-R diagram}
\label{sec:hr}

The atmospheric parameters obtained in this work are necessary ingredients for seismic modelling.
Our sample of A and F stars contains pulsating $\delta$\,Sct, $\gamma$\,Dor, and hybrid stars.
To find out if the atmospheric parameters of the analysed stars are consistent with theoretical instability strips in this part of the H-R diagram,
we performed an analysis of the \kepler\ data for all objects investigated here.
We used the multi-scale MAP light curves available through the MAST\footnote{https://archive.stsci.edu/kepler/} database.
We considered all the long cadence data available for a given star. If necessary, short cadence data were used to verify our classification.
We visually examined all light curves, the frequency spectra, and detected frequencies to identify candidate $\delta$\,Sct, $\gamma$\,Dor, hybrid, and other types of variability.
The results of our initial classification are presented in the last column of Table\,\ref{table1}. 

Most of the stars were classified as hybrid $\delta$\,Sct/$\gamma$\,Dor variables ($29$), including two stars showing HADS pulsations with possible $\gamma$\,Dor frequencies
(KIC\,3868420 and KIC\,9408694). In this work all objects for which we can find frequencies typical for $\delta$\,Sct and $\gamma$\,Dor stars are termed ``hybrid''.
We classified $8$ stars as $\delta$\,Sct (single stars and binary systems), and $4$ as pure $\gamma$\,Dor variables.
For $4$ stars the classification was not clear, so we denoted these stars as ``unknown'' variables.
Three of these stars have early spectral types from A3 to A6, all with luminosity class V (KIC\,5880360, KIC\,8211500, and KIC\,11622328)
but they show variability with frequencies typical for $\gamma$\,Dor stars.
The other ``unknown'' is the F6\,III star, KIC\,10341072. Two stars were classified as constant: KIC\,9349245 and KIC\,9941662.
The latter has planetary transits \citep[\textit{Kepler}-13,][]{2014ApJ...788...92S}.
Solar like oscillations were discovered for the three stars with the lowest temperature: KIC\,9410862, KIC\,8694723
(metal weak star hF6\,mF2\,gF5\,V), and KIC\,6370489. Moreover, we found $6$ pulsating stars in binary systems,
including eclipsing binaries (KIC\,2162283, KIC\,3858884, and KIC\,10661783). 

All of the investigated stars are shown in the \teff\ $-$ \logg\ diagram (Fig.\,\ref{hrall}).
Evolutionary tracks, calculated with Time Dependent Convection \citep[TDC,][]{2005A&A...434.1055G},
for stars having [Fe/H]\,$=0.0$ and $\alpha_{\rm MLT} = 1.8$ covering a range of masses from $1.4$ to $2.8$\,M$_\odot$ are also shown.
The instability strips for $\delta$\,Sct (solid line) and $\gamma$\,Dor (dashed line) are shown as well. 
As can be seen in Fig.\,\ref{hrall}, all the Am stars are located firmly inside the instability strip.
Only one Am star was classified as a non-pulsating star (KIC\,9941662),two are $\delta$\,Sct stars (KIC\,2162283 and KIC\,3858884), one is $\gamma$\,Dor (KIC\,7767565),
and one is possible hybrid (KIC\,3868420). 
As for the $\lambda$\,Boo stars, one is located inside the $\delta$\,Sct instability strip (KIC\,11973705) and one is hotter (KIC\,9828226).
All $\lambda$\,Boo stars were classified as $\delta$\,Sct/$\gamma$\,Dor variables.
Two chemically normal stars are located well outside the blue edge of the $\delta$\,Sct instability strip,
KIC\,11044547, classified as possible hybrid with clear $\delta$\,Sct and possible $\gamma$\,Dor frequencies,
and KIC\,9650390 classified as a clear hybrid star. Their temperatures are too high for hybrid $\delta$\,Sct/$\gamma$\,Dor stars.
These two stars need further investigation to verify the possibility of contamination by another star.
The other hybrid stars are located inside the $\delta$\,Sct instability strip. 

%&&&&&&&&&&&&&&&&&&&&&&&&&&&&&&&&&&&&&&&&&&&&&&&&&&&&&&&&&&&&&&&&&&&&&&&&&&&&&&&&&&&&&&&&&&&&&&&&&&&&&&&&&&&&&&&&&&

\section{Conclusions}
\label{sec:conclusions}

We have analysed FIES/NOT spectra of $50$ A and F stars observed by \kepler. 
Spectral classification showed that our sample consists mostly of non-evolved stars of luminosity types V to IV and revealed $9$ CP stars of Am and $\lambda$\,Bo\"{o}tis types.
Six of them were discovered as new CP stars. Atmospheric peculiarities of these stars were confirmed by detailed investigation of their stellar spectra. 

We determined atmospheric parameters -- \teff, \logg, \micro, \vsini\ and chemical abundances -- using spectral synthesis.
Initial values of \teff\ and \logg\ were derived from \citep{huber2014} (mostly photometric values), SED, and Balmer lines analysis.
For discussion of the parameters we used an enlarged sample of stars, including those analysed in Paper\,I, using the same method. The combined sample consists of $167$ stars. 
We have confirmed that atmospheric parameters obtained from photometry (e.g.\ H2014), particularly \logg, are markedly inferior to those obtained spectroscopically from iron lines.
The ability of photometry to derive these quickly for a large number of stars means this method will continue to have a role in input catalogues for space telescopes,
but spectroscopy should be utilised when high accuracy and precision is required.

We have mapped the change in microturbulent velocities as a function of \teff\ for normal stars and for Am stars in a self-consistent manner,
and buck the trend in finding no microturbulence excess for Am stars over normal stars.
We confirm suspicions that earlier results of that nature \citep[e.g.][]{landstreet2009}
failed to account for the increase in microturbulence for {\it all} stars in the $7000-8000$\,K temperature range where Am stars are predominantly found \citep{murphy2014}.
We also find no correlation between \micro\ and \vsini.
The narrow \logg\ range among our targets prevented a similar analysis for that parameter.

The stars analysed here have \vsini\ values ranging from $4$ to about $280$\,\kms.
There are seven objects with $v\sin i < 30$\,\kms\ in our sample, including CP and non-CP stars.
The mean value of \vsini\ for the whole sample is $125$\,\kms, and the median is $110$\,\kms, consistent with literature results for A/F stars in general \citep[e.g.][]{royer2009}.
Abundances are hard to measure for rapidly rotating A stars, but we confirm that they show large star-to-star variations \citep{hill&landstreet1993},
especially where \vsini\ exceeds $150$\,\kms.

Homogeneous analysis of large samples of A and F stars frames the discussion of the nature of their atmospheres, e.g. the \vsini\ distribution
and the influence of fast rotation on chemical abundances, and the dependence of microturbulent velocity on effective temperature and surface gravity.
This provides a platform for greater understanding of atomic diffusion, mixing processes and angular momentum transport, which are extremely important but remain ill-understood.
In addition, the provision of reliable stellar atmospheric parameters from spectroscopy is crucial to successful asteroseismic investigation of these objects,
extending analyses from the line-forming region analysed spectroscopically to the deep interior of A/F stars probed by g\:modes \citep{kurtzetal2014,saioetal2015,murphyetal2016a,schmid&aerts2016}.

In the next papers we enlarge our sample for the other A-F stars from the \kepler\ field (Poli\'nska et al., in preparation)
as well as \textit{K2} fields (Niemczura et al., in preparation). Both samples consist of possible $\delta$\,Sct, $\gamma$\,Dor and hybrid stars.
All the stars will be analysed with the same methods, using the same atmospheric models, and atomic data.
The enlarged sample will consist of about $300$ stars, what let us to observationally define instability regions for A-F pulsating stars.
In addition, we check the possible correlations of pulsational characteristics of stars, obtained from \kepler\ data analysis with their atmospheric parameters
and projected rotational velocity.

%----------

%&&&&&&&&&&&&&&&&&&&&&&&&&&&&&&&&&&&&&&&&&&&&&&&&&&&&&&&&&&&&&&&&&&&&&&&&&&&&&&&&&&&&&&&&&&&&&&&&&&&&&&&&&&&&&&&&&&

\section*{Acknowledgments}
EN and MP acknowledge the Polish National Science Center grant no. 2014/13/B/ST9/00902. 
Calculations have been carried out at the Wroc{\l}aw Centre for Networking and Supercomputing (http://www.wcss.pl), grant No.\,214.
This paper includes data collected by the \kepler\ mission. Funding for the \kepler\ mission is provided by the NASA Science Mission directorate.

%&&&&&&&&&&&&&&&&&&&&&&&&&&&&&&&&&&&&&&&&&&&&&&&&&&&&&&&&&&&&&&&&&&&&&&&&&&&&&&&&&&&&&&&&&&&&&&&&&&&&&&&&&&&&&&&&&&

%%%%%%%%%%%%%%%%%%%%%%%%%%%%%%%%%%%%%%%%%%%%%%%%%%%%%%%%%%%%%%%%%%%%%%%%%%%%%%%%%%%%%%%%%%%%%%%%%%%%%%%%%%%%%

\appendix

\section{Comparison with the literature}

Most stars presented in this work are analysed here for the first time. We summarise the literature on the remainder in this appendix.

\begin{itemize}
\item KIC\,2162283 -- eclipsing binary system. The star was analysed here as a single object and the results of all methods are consistent.
\citet{2014MNRAS.437.3473A} combined the data from the photometric surveys to construct SED for this binary system and obtained primary and secondary stellar temperatures
of $7530\pm350$ and $7520\pm540$\,K.

\item KIC\,3219256 -- our results are consistent with each other. \citet{2011MNRAS.411.1167C}
used low resolution data to obtain similar effective temperature $7500\pm150$\,K and surface gravity $3.6\pm0.1$\,dex.

\item KIC\,3429637 -- the atmospheric parameters obtained in this work are consistent.
The effective temperatures from SED with TD1 ($7340$\,K) and without TD1 data ($7260$\,K) are also similar.
\citet{2009A&A...498..961R} mentioned this star in the catalogue of chemically peculiar objects, without pointing the peculiarity type.
\citet{2011MNRAS.411.1167C} analysed medium resolution spectra and obtained effective temperature $7100\pm150$\,K, similar to our value.
The surface gravity determined by them, $3.00\pm0.20$\,dex is lower that our value, $3.40\pm0.10$\,dex.
\citet{2012MNRAS.427..343M} determined fundamental atmospheric parameters from the analysis of the high-resolution spectra.
The atmospheric parameters obtained by them, effective temperature $7300\pm100$\,K, surface gravity $3.0\pm0.1$\,dex, microturbulence $4.0\pm0.5$\,\kms,
and projected rotational velocity $51\pm1$\,\kms\ are close to our results. They also confirmed earlier characterizations of KIC\,3429637 as a marginal Am star. 

\item KIC\,3453494 -- our results are consistent. The star was analysed by \citet{2012MNRAS.422.2960T}.
They found similar effective temperature from SED analysis, $7840\pm240$\,K, and from spectral analysis, $7737\pm57$\,K.
Also their \logg\ value is consistent with our result. On the other hand, the microturbulence value obtained here,
$2.00\pm0.5$\,\kms is lower than $3.24\pm0.66$\,\kms determined by \citet{2012MNRAS.422.2960T}.
It is the result of very fast rotation of this star, which makes it difficult to derive the correct value of microturbulence.

\item KIC\,3868420 -- this star belongs to a multiple system consisting of two brighter and closer components, $10.35$ and $10.61$\,mag, with separation 2.5",
plus a wide (26") third component with $V=11.71\,mag$. For this reason the photometric magnitudes used for SED construction are unreliable,
as they were obtained for the combined light of the close pair. There is insufficient photometry of the individual stars to obtain SED fits for each star.
Moreover, this is a HADS pulsator, so photometry will sample the star at different phases and brightnesses.
Our results from Balmer and iron lines analysis are consistent.
\citet{2013ApJ...773..181N} defined this star as HADS with \teff\,$=7612$\,K, \logg\,$=3.71$, \micro\,$=2.3$\,\kms, macroturbulence $23.0$\,\kms,
and \vsini\,$=5.7$\,\kms\ from spectrum synthesis.
Their atmospheric parameters are consistent with our results except for microturbulence, which is lower.
The differences in \vsini\ are caused by the inclusion of macroturbulence.  

\item KIC\,5988140 -- results obtained here are consistent with each other and with Paper\,I. \citet{2011MNRAS.411.1167C} found a lower \teff, $7400\pm150$\,K,
but the same \logg\ $3.7\pm0.3$ from the analysis of low resolution spectroscopy.
\citet{2013A&A...549A.104L} found effective temperature $7600\pm30$\,K, surface gravity $3.40\pm0.12$\,dex, projected rotation velocity $52\pm1.5$\,\kms,
and microturbulence $3.16\pm0.20$\,\kms\ from high-resolution spectroscopy and spectrum synthesis. 

\item KIC\,5880360 -- the atmospheric parameters obtained in this work are consistent. 
\citet{2012MNRAS.427..343M} determined effective temperature $7510$\,K from the SED analysis, significantly lower than the result obtained by us ($8070\pm220$\,K).

\item KIC\,6370489 -- our results are consistent. \citet{2013MNRAS.434.1422M} determined similar effective temperature $6241\pm116$\,K and surface gravity $3.98\pm0.21$\,dex
from the analysis of high-resolution spectra. Also \citet{2014ApJS..210....1C} obtained similar effective temperatures of $6300\pm80$\,K and $6324\pm144$\,K from photometric data.

\item KIC\,7119530 -- the results from iron lines analysis are consistent with Paper\,I.
\citet{2011MNRAS.411.1167C} determined a lower effective temperature, $7500$\,K, from low resolution spectroscopy and Balmer-line analysis.
Their surface gravity, $3.6\pm0.3$\,dex is consistent with ours.

\item KIC\,7748238 -- the atmospheric parameters obtained in this work agree very well with each other.
\citet{2012MNRAS.422.2960T} found similar values of \teff\ ($7264$\,K), \logg\ ($3.96$) and \micro\ ($3.53$\,\kms) from high-resolution spectrum synthesis.

\item KIC\,7798339 -- our results are consistent. \citet{1997A&AS..126...21N} analysed echelle spectra in the narrow wavelength range $5165.77$--$5211.25$\,\AA\ to
obtain $T_{\rm eff} = 7000$\,K, $\log g = 3.5$ and $v\sin i = 15.4$\,\kms.
These values are in good agreement with ours. \citet{2011MNRAS.411.1167C} analysed medium resolution spectra and obtained effective temperature $6700\pm150$\,K
and surface gravity $3.7\pm0.3$\,dex, close to our results. \citet{2012MNRAS.427..343M} determined $T_{\rm eff}=6900$\,K from SED analysis, consistent with ours.

\item KIC\,8211500 -- the atmospheric parameters obtained in this work are consistent.
It is a binary system WDS18462+4408 of stars with magnitudes $8.19$ and $11.28$ and separation 0.7".
The SED result is for combined light, but the large magnitude difference means that the faint companion should not affect the results.
\citet{2012MNRAS.427..343M} determined $T_{\rm eff}=7709$\,K in accordance with our results.
Similarly, \citet{2015MNRAS.447.3948M} gave $T_{\rm eff} = 7800$\,K and $\log g = 3.8$. 

\item KIC\,8694723 -- our results are consistent. Atmospheric parameters of this star determined from the analysis of high-resolution spectra were available before.
\citet{2013MNRAS.434.1422M} obtained \teff\ and \logg\ ($6287\pm116$\,K, $4.00\pm0.21$), iron abundance ($-0.38\pm0.22$),
and \vsini\ value ($3.8\pm0.7$\,\kms) consistent with our results.
\citet{2013MNRAS.433.3227K} derived $T_{\rm eff} = 6287\pm60$\,K, $\log g = 4,079\pm0.035$, iron abundance $-0.59\pm0.06$,
and projected rotational velocity $6.6$\,\kms\ in accordance with our values.
The atmospheric parameters and projected rotational velocities close to the values obtained here were also given by \citet{2012MNRAS.423..122B}.
Higher effective temperatures (about $6400-6500$\,K) were determined by \citet{2014A&A...572A..95M} from the equivalent-width method and high-resolution spectroscopy.

\item KIC\,9349245 -- the atmospheric parameter indicators adopted in this work agree very well with each other.
\citet{2009A&A...498..961R} included this star in their catalogue of Ap, HgMn and Am stars, but without the classification of the peculiarity type.
\citet{2013MNRAS.431.2240B} classified this star as rotationally variable star on the basis of \kepler\ data.
\citet{2014MNRAS.441.3543B} determined the effective temperature of the star as $7690$\,K, lower than our results.
\citet{2015MNRAS.451..184C} analysed the SED and high-resolution spectra and determined the atmospheric parameters of this star.
The results from the SED ($8250\pm250$\,K) and from spectrum synthesis ($T_{\rm eff} = 8300\pm200$\,K, $\log g = 4.00\pm0.30$, $\xi_{\rm t} = 3.1\pm0.5$\,\kms,
and \vsini\ $80\pm3$\,\kms) are consistent with the results of our analysis.

\item KIC\,9408694 (V2367\,Cyg) -- is a HADS pulsator, so photometry used for SED construction can sample the star at different phases and brightnesses.
Our results are consistent with each other, and in agreement with the valued determined before.
\citet{2012MNRAS.419.3028B} analysed low-resolution spectra and obtained \teff\ $7300\pm150$\,K, \logg\ $3.5\pm0.1$ consistent with our results.
Also \citet{2013MNRAS.428.3551U} analysed low-resolution data to obtain similar values.
The best fit to the H$\beta$ and H$\alpha$ lines was obtained for effective temperature $7250$\,K and surface gravity $3.5$\,dex.

\item KIC\,9410862 -- our results are consistent.
\citet{2014ApJS..210....1C} determined effective temperatures from SDSS ($6230\pm53$\,K) and IRFM ($6143\pm148$\,K) photometry, in agreement with our results.

\item KIC\,10030943 -- this is an eclipsing binary. The atmospheric parameter indicators adopted in this work agree with each other,
except for the SED \teff. This may be the result of the influence of the secondary star on the photometric indices.
Using photometric methods,
\citet{2014MNRAS.437.3473A} determined atmospheric parameters of both objects, effective temperatures $6761\pm354$\,K and $6727\pm549$\,K, respectively.

\item KIC\,10661783 -- our results are consistent. This is an eclipsing binary of $\beta$\,Lyr type (semi-detached).
It was analysed by \citet{2011MNRAS.414.2413S} as an eclipsing binary system with $\delta$\,Sct pulsations.
They determined temperatures of both stars at $8000\pm160$ and $6500$\,K.
\citet{2013A&A...557A..79L} analysed physical properties of both stars and found effective temperatures ($7764\pm54$, $5980\pm72$\,K),
surface gravities ($3.9$, $3.6$\,dex), and projected rotational velocities ($79\pm4$, $48\pm3$\,\kms).
Our results are consistent with the atmospheric parameters of the primary component.
In a catalogue of \kepler\ eclipsing binary stars, \citet{2014MNRAS.437.3473A} give wildly discordant effective temperatures of $9197\pm475$\,K and $7714\pm808$\,K
for primary and secondary, respectively.

\item KIC\,11973705 -- is a binary star, probably consisting of a late-B and a late-A star. We noted ellipsoidal variability in the \kepler\ light curve.
The atmospheric parameters obtained in this work are consistent with each other, and correspond to the A star.
The effective temperature from H2014 differs significantly from the other results, at $11000$\,K.

\citet{2011A&A...526A.124L} performed spectral analysis of \kepler\ SPB and $\beta$\,Cep candidate stars.
They classified this star as SPB with spectral type B8.5\,V-IV (HD\,234999) and determined \teff\ from SED analysis ($7920\pm100$\,K)
taking into account TD-1 data. The effective temperature obtained by \citet{2011A&A...526A.124L} from spectral analysis is similar to the one in H2014, $11150$\,K.
They therefore appear to have sampled both stars.

\citet{2011MNRAS.411.1167C} noted that the spectral type recorded in the Henry Draper catalogue (B9),
is an entire spectral class too early in comparison with their classification of A9.5\,V.
They classified the star as a $\delta$\,Sct variable in a binary system.
Their atmospheric parameters obtained from the analysis of low-resolution (R~$5000$)
spectroscopy and 2MASS, uvby$\beta$ and IRFM photometry are in accordance with our results.

\citet{2011MNRAS.413.2403B} searched \kepler\ observations for variability seen in B stars.
The star was classified as binary system with an SPB and a $\delta$\,Sct component, with spectral type B8.5\,V-IVb.
In a catalogue of effective temperatures for \kepler\ eclipsing binary stars, \citet{2014MNRAS.437.3473A} gave $7918\pm356$\,K and $6528\pm556$\,K,
not particularly capturing the nature of either star. \citet{2015MNRAS.451.1445B} classified this star as Maia variable. 

\end{itemize}

\label{lastpage}

%&&&&&&&&&&&&&&&&&&&&&&&&&&&&&&&&&&&&&&&&&&&&&&&&&&&&&&&&&&&&&&&&&&&&&&&&&&&&&&&&&&&&&&&&&&&&&&&&&&&&&&&&&&&&&&&&&&

\section{Additional tables}

\begin{landscape}

\begin{table}
\begin{scriptsize}
\label{abundance-table}
\caption{Abundances of chemical elements for the analysed stars (on the scale in which $\log \epsilon$(H) = $12$).
The median value is given; standard deviations were calculated only if the number of analysed parts or lines (depending on \vsini\ of the star) is greater than three.
In other cases the average value calculated from standard deviations of other elements are given.
Number of analysed parts or lines is given in brackets.}

\begin{tabular}{lllllllllll}

\midrule
   & KIC\,10030943 &     KIC\,10341072 &     KIC\,10661783 &     KIC\,10717871 &     KIC\,11044547 &     KIC\,11622328 &     KIC\,11973705 &      KIC\,2162283 &      KIC\,2694337 &      KIC\,3219256\\
    C & 8.65$\pm$0.05 (6) &   8.75$\pm$0.10 (5) &   8.17$\pm$0.20 (4) &   8.41$\pm$0.25 (5) &   8.49$\pm$0.35 (5) &   8.43$\pm$0.14 (3) &   8.45$\pm$0.19 (2) &  8.64$\pm$0.37 (14) &  8.58$\pm$0.19 (10) &  8.32$\pm$0.11 (2) \\
    N               & $-$ &                 $-$ &   8.42$\pm$0.20 (1) &                 $-$ &   8.28$\pm$0.18 (1) &                 $-$ &                 $-$ &   8.72$\pm$0.19 (2) &   8.34$\pm$0.17 (1) &                $-$ \\
    O & 8.93$\pm$0.15 (1) &                 $-$ &   8.92$\pm$0.20 (2) &   9.02$\pm$0.19 (1) &   8.92$\pm$0.18 (2) &   8.93$\pm$0.13 (2) &   8.49$\pm$0.19 (2) &   9.17$\pm$0.19 (2) &   9.02$\pm$0.17 (1) &  8.78$\pm$0.11 (1) \\
   Ne               & $-$ &                 $-$ &                 $-$ &                 $-$ &                 $-$ &                 $-$ &                 $-$ &                 $-$ &                 $-$ &                $-$ \\
   Na & 5.98$\pm$0.14 (3) &   7.06$\pm$0.09 (3) &   6.58$\pm$0.20 (2) &   6.44$\pm$0.19 (2) &   6.80$\pm$0.18 (1) &                 $-$ &                 $-$ &   6.61$\pm$0.13 (3) &   6.70$\pm$0.17 (2) &  6.59$\pm$0.11 (1) \\
   Mg & 7.30$\pm$0.23 (6) &   7.74$\pm$0.23 (7) &   7.79$\pm$0.18 (5) &   7.65$\pm$0.12 (5) &   7.68$\pm$0.14 (8) &   7.71$\pm$0.05 (4) &   6.81$\pm$0.11 (3) &   7.63$\pm$0.17 (8) &  7.80$\pm$0.12 (12) &  7.66$\pm$0.13 (4) \\
   Al               & $-$ &   6.73$\pm$0.18 (1) &                 $-$ &                 $-$ &   6.30$\pm$0.18 (1) &                 $-$ &                 $-$ &   6.78$\pm$0.19 (1) &                 $-$ &                $-$ \\
   Si& 7.19$\pm$0.24 (12) &  7.69$\pm$0.20 (13) &  7.51$\pm$0.32 (11) &  7.52$\pm$0.14 (10) &   7.74$\pm$0.15 (8) &   7.61$\pm$0.39 (3) &   7.25$\pm$0.37 (3) &  7.65$\pm$0.25 (36) &  7.55$\pm$0.20 (17) &  7.82$\pm$0.17 (7) \\
    P               & $-$ &                 $-$ &                 $-$ &                 $-$ &                 $-$ &                 $-$ &                 $-$ &                 $-$ &                 $-$ &                $-$ \\
    S & 7.55$\pm$0.15 (2) &   7.75$\pm$0.18 (2) &   7.70$\pm$0.20 (2) &   7.14$\pm$0.19 (2) &                 $-$ &                 $-$ &                 $-$ &  7.61$\pm$0.16 (11) &   7.39$\pm$0.23 (3) &  7.05$\pm$0.11 (1) \\
   Cl               & $-$ &                 $-$ &                 $-$ &                 $-$ &                 $-$ &                 $-$ &                 $-$ &                 $-$ &                 $-$ &                $-$ \\
    K               & $-$ &                 $-$ &                 $-$ &                 $-$ &                 $-$ &                 $-$ &                 $-$ &                 $-$ &                 $-$ &                $-$ \\
   Ca& 6.18$\pm$0.17 (15) &  6.80$\pm$0.25 (18) &  6.33$\pm$0.25 (13) &  6.70$\pm$0.22 (12) &  6.78$\pm$0.16 (15) &   6.33$\pm$0.10 (5) &   5.48$\pm$0.18 (8) &  6.48$\pm$0.21 (22) &  6.68$\pm$0.15 (13) &  6.43$\pm$0.11 (9) \\
   Sc & 2.94$\pm$0.12 (5) &   3.64$\pm$0.15 (9) &   3.36$\pm$0.19 (6) &   3.50$\pm$0.39 (5) &   3.37$\pm$0.24 (6) &   3.26$\pm$0.04 (4) &   2.11$\pm$0.25 (3) &   3.04$\pm$0.23 (9) &   3.21$\pm$0.14 (9) &  3.13$\pm$0.08 (5) \\
   Ti& 4.92$\pm$0.11 (25) &  5.37$\pm$0.22 (25) &  5.18$\pm$0.23 (21) &  5.12$\pm$0.26 (17) &  5.20$\pm$0.16 (27) &  5.14$\pm$0.09 (10) &  4.03$\pm$0.17 (10) &  5.26$\pm$0.20 (50) &  5.13$\pm$0.18 (23) & 5.07$\pm$0.14 (15) \\
    V & 4.23$\pm$0.21 (3) &   4.39$\pm$0.35 (4) &   4.10$\pm$0.18 (4) &   4.55$\pm$0.19 (2) &   4.87$\pm$0.18 (2) &   4.22$\pm$0.13 (1) &                 $-$ &   4.89$\pm$0.25 (9) &   3.96$\pm$0.27 (3) &                $-$ \\
   Cr& 5.42$\pm$0.16 (20) &  5.83$\pm$0.21 (23) &  5.80$\pm$0.18 (18) &  5.59$\pm$0.20 (16) &  5.70$\pm$0.11 (16) &  5.67$\pm$0.12 (10) &   4.66$\pm$0.17 (5) &  6.08$\pm$0.19 (64) &  5.71$\pm$0.17 (23) & 5.57$\pm$0.12 (12) \\
   Mn & 5.30$\pm$0.24 (4) &   5.80$\pm$0.25 (6) &   5.15$\pm$0.29 (6) &   5.80$\pm$0.09 (3) &   5.48$\pm$0.18 (2) &   5.29$\pm$0.13 (1) &                 $-$ &  5.47$\pm$0.25 (17) &   5.34$\pm$0.13 (5) &  5.07$\pm$0.11 (2) \\
   Fe& 7.07$\pm$0.10 (58) &  7.63$\pm$0.08 (45) &  7.51$\pm$0.12 (47) &  7.45$\pm$0.11 (45) &  7.63$\pm$0.13 (63) &  7.40$\pm$0.09 (27) &  6.50$\pm$0.14 (24) & 7.71$\pm$0.14 (187) &  7.60$\pm$0.07 (59) & 7.40$\pm$0.10 (45) \\
   Co               & $-$ &   5.96$\pm$0.18 (2) &                 $-$ &   5.78$\pm$0.19 (1) &                 $-$ &                 $-$ &                 $-$ &   5.59$\pm$0.31 (4) &   5.06$\pm$0.17 (1) &                $-$ \\
   Ni& 5.95$\pm$0.14 (21) &  6.53$\pm$0.21 (29) &  6.22$\pm$0.20 (13) &  6.22$\pm$0.14 (11) &   6.61$\pm$0.23 (8) &   6.55$\pm$0.22 (3) &   5.67$\pm$0.20 (5) &  6.65$\pm$0.18 (72) &  6.36$\pm$0.21 (25) &  6.23$\pm$0.10 (9) \\
   Cu & 4.45$\pm$0.15 (2) &   4.30$\pm$0.09 (3) &   3.36$\pm$0.20 (1) &   4.06$\pm$0.19 (1) &                 $-$ &                 $-$ &                 $-$ &   4.54$\pm$0.12 (3) &   3.98$\pm$0.17 (2) &                $-$ \\
   Zn & 4.39$\pm$0.15 (1) &   4.19$\pm$0.18 (1) &                 $-$ &   4.21$\pm$0.19 (1) &                 $-$ &                 $-$ &                 $-$ &   4.88$\pm$0.19 (2) &   3.96$\pm$0.17 (1) &                $-$ \\
   Ga               & $-$ &                 $-$ &                 $-$ &                 $-$ &                 $-$ &                 $-$ &                 $-$ &   4.02$\pm$0.19 (1) &                 $-$ &                $-$ \\
   Sr & 3.08$\pm$0.15 (1) &   3.52$\pm$0.18 (2) &   3.43$\pm$0.20 (1) &   2.98$\pm$0.19 (1) &   3.11$\pm$0.18 (1) &   2.31$\pm$0.13 (1) &   2.02$\pm$0.19 (1) &   3.94$\pm$0.01 (3) &   3.39$\pm$0.17 (2) &  3.26$\pm$0.11 (1) \\
    Y & 2.34$\pm$0.18 (6) &   2.76$\pm$0.26 (5) &   2.32$\pm$0.12 (4) &   2.47$\pm$0.20 (3) &   2.19$\pm$0.18 (2) &   1.82$\pm$0.13 (1) &   1.97$\pm$0.19 (1) &  3.02$\pm$0.20 (15) &   2.49$\pm$0.21 (5) &                $-$ \\
   Zr & 2.87$\pm$0.07 (4) &   3.12$\pm$0.01 (3) &   2.78$\pm$0.21 (3) &   2.65$\pm$0.23 (4) &   3.90$\pm$0.18 (1) &                 $-$ &                 $-$ &  3.26$\pm$0.26 (10) &   3.05$\pm$0.13 (5) &                $-$ \\
   Ba & 2.59$\pm$0.13 (3) &   2.90$\pm$0.18 (2) &   3.26$\pm$0.20 (2) &   2.80$\pm$0.14 (3) &   2.32$\pm$0.18 (2) &   2.77$\pm$0.13 (1) &   1.66$\pm$0.19 (2) &   3.64$\pm$0.19 (4) &   2.83$\pm$0.17 (2) &  3.05$\pm$0.11 (2) \\
   La & 1.39$\pm$0.15 (2) &   1.35$\pm$0.18 (1) &   1.56$\pm$0.20 (1) &   1.61$\pm$0.19 (1) &                 $-$ &                 $-$ &                 $-$ &   2.55$\pm$0.10 (5) &   1.73$\pm$0.17 (1) &                $-$ \\
   Ce & 1.85$\pm$0.15 (2) &                 $-$ &                 $-$ &                 $-$ &                 $-$ &                 $-$ &                 $-$ &  2.85$\pm$0.14 (12) &   1.91$\pm$0.17 (1) &                $-$ \\
   Pr               & $-$ &                 $-$ &                 $-$ &                 $-$ &                 $-$ &                 $-$ &                 $-$ &                 $-$ &                 $-$ &                $-$ \\
   Nd & 2.08$\pm$0.15 (2) &   2.38$\pm$0.18 (2) &                 $-$ &                 $-$ &                 $-$ &                 $-$ &                 $-$ &   2.43$\pm$0.16 (9) &   2.37$\pm$0.17 (2) &                $-$ \\
   Sm               & $-$ &                 $-$ &                 $-$ &                 $-$ &                 $-$ &                 $-$ &                 $-$ &                 $-$ &                 $-$ &                $-$ \\
   Eu               & $-$ &   2.49$\pm$0.18 (1) &                 $-$ &                 $-$ &                 $-$ &                 $-$ &                 $-$ &                 $-$ &                 $-$ &                $-$ \\
   Gd               & $-$ &                 $-$ &                 $-$ &                 $-$ &                 $-$ &                 $-$ &                 $-$ &                 $-$ &                 $-$ &                $-$ \\
   Dy               & $-$ &                 $-$ &                 $-$ &                 $-$ &                 $-$ &                 $-$ &                 $-$ &                 $-$ &                 $-$ &                $-$ \\
   Er               & $-$ &                 $-$ &                 $-$ &                 $-$ &                 $-$ &                 $-$ &                 $-$ &                 $-$ &                 $-$ &                $-$ \\

\bottomrule
\end{tabular}
\end{scriptsize}
\end{table}
\end{landscape}

\newpage

\setcounter{table}{0}

\begin{landscape}

\begin{table}
\begin{scriptsize}
\caption{continuation}
\begin{tabular}{lllllllllll}
\midrule
           & KIC3331147\, &      KIC\,3429637 &      KIC\,3453494 &      KIC\,3868420 &      KIC\,4647763 &      KIC\,5880360 &      KIC\,5988140 &      KIC\,6123324 &      KIC\,6279848 &      KIC\,6370489\\
    C& 8.36$\pm$0.12 (12) &   8.21$\pm$0.14 (8) &   7.97$\pm$0.11 (2) &  7.95$\pm$0.30 (11) &   8.31$\pm$0.08 (5) &   8.27$\pm$0.16 (1) &   8.33$\pm$0.20 (7) &   8.94$\pm$0.18 (2) &  8.08$\pm$0.23 (13) & 8.21$\pm$0.22 (24) \\
    N               & $-$ &                 $-$ &                 $-$ &                 $-$ &                 $-$ &                 $-$ &                 $-$ &                 $-$ &                 $-$ &                $-$ \\
    O & 9.15$\pm$0.17 (1) &   8.39$\pm$0.13 (1) &   8.43$\pm$0.11 (1) &   8.49$\pm$0.16 (1) &                 $-$ &                 $-$ &   8.72$\pm$0.16 (1) &                 $-$ &   8.19$\pm$0.17 (1) &  8.52$\pm$0.22 (1) \\
   Ne               & $-$ &                 $-$ &                 $-$ &                 $-$ &                 $-$ &                 $-$ &                 $-$ &                 $-$ &                 $-$ &                $-$ \\
   Na & 6.52$\pm$0.23 (3) &   6.57$\pm$0.13 (2) &                 $-$ &   6.67$\pm$0.34 (6) &   6.64$\pm$0.26 (4) &                 $-$ &   5.32$\pm$0.16 (1) &                 $-$ &   6.37$\pm$0.23 (3) & 5.95$\pm$0.13 (10) \\
   Mg & 7.79$\pm$0.04 (6) &   7.77$\pm$0.13 (5) &   8.02$\pm$0.01 (3) &  7.42$\pm$0.23 (13) &   7.47$\pm$0.10 (7) &   7.23$\pm$0.35 (4) &   7.80$\pm$0.05 (4) &   7.43$\pm$0.18 (2) &  7.36$\pm$0.18 (10) & 7.37$\pm$0.11 (12) \\
   Al               & $-$ &                 $-$ &                 $-$ &   6.67$\pm$0.16 (1) &                 $-$ &                 $-$ &                 $-$ &                 $-$ &                 $-$ &                $-$ \\
   Si& 7.60$\pm$0.19 (25) &  7.47$\pm$0.17 (11) &   7.98$\pm$0.16 (4) &  7.68$\pm$0.25 (24) &  7.48$\pm$0.31 (12) &   7.05$\pm$0.16 (2) &  7.62$\pm$0.26 (11) &   7.50$\pm$0.18 (2) &  7.36$\pm$0.25 (19) & 7.29$\pm$0.26 (53) \\
    P               & $-$ &                 $-$ &                 $-$ &                 $-$ &                 $-$ &                 $-$ &                 $-$ &                 $-$ &                 $-$ &                $-$ \\
    S & 7.28$\pm$0.15 (4) &   7.45$\pm$0.08 (4) &                 $-$ &  7.14$\pm$0.24 (10) &   7.29$\pm$0.15 (1) &                 $-$ &   7.40$\pm$0.16 (2) &                 $-$ &   7.42$\pm$0.04 (3) & 7.01$\pm$0.14 (10) \\
   Cl               & $-$ &                 $-$ &                 $-$ &                 $-$ &                 $-$ &                 $-$ &                 $-$ &                 $-$ &                 $-$ &                $-$ \\
    K               & $-$ &                 $-$ &                 $-$ &                 $-$ &                 $-$ &                 $-$ &                 $-$ &                 $-$ &                 $-$ &                $-$ \\
   Ca& 6.58$\pm$0.26 (20) &  6.70$\pm$0.12 (17) &   6.46$\pm$0.09 (3) &  5.66$\pm$0.21 (22) &  6.60$\pm$0.14 (14) &   6.20$\pm$0.15 (6) &  6.53$\pm$0.20 (22) &   5.60$\pm$0.25 (5) &  6.37$\pm$0.12 (22) & 6.11$\pm$0.17 (49) \\
   Sc & 2.91$\pm$0.06 (7) &  3.33$\pm$0.07 (11) &   2.95$\pm$0.11 (2) &   2.59$\pm$0.09 (9) &   3.30$\pm$0.14 (7) &   2.69$\pm$0.19 (4) &   3.18$\pm$0.10 (9) &   2.70$\pm$0.20 (5) &  3.14$\pm$0.09 (10) & 2.93$\pm$0.17 (21) \\
   Ti& 4.92$\pm$0.15 (31) &  5.03$\pm$0.18 (36) &   4.99$\pm$0.19 (9) &  4.93$\pm$0.19 (67) &  4.93$\pm$0.12 (35) &   4.61$\pm$0.10 (7) &  5.13$\pm$0.13 (28) &  4.58$\pm$0.23 (11) &  4.83$\pm$0.13 (52) &4.73$\pm$0.19 (180) \\
    V & 4.37$\pm$0.17 (2) &   4.50$\pm$0.15 (7) &                 $-$ &  4.66$\pm$0.17 (10) &   4.18$\pm$0.15 (2) &                 $-$ &   4.61$\pm$0.16 (2) &                 $-$ &   4.32$\pm$0.29 (8) & 3.73$\pm$0.19 (29) \\
   Cr& 5.60$\pm$0.11 (29) &  5.76$\pm$0.14 (34) &   5.47$\pm$0.12 (3) &  6.23$\pm$0.15 (94) &  5.58$\pm$0.14 (25) &   5.53$\pm$0.15 (7) &  5.66$\pm$0.18 (25) &   5.08$\pm$0.15 (5) &  5.50$\pm$0.17 (55) &5.30$\pm$0.20 (155) \\
   Mn & 5.32$\pm$0.25 (6) &   5.46$\pm$0.16 (8) &   5.30$\pm$0.11 (1) &  5.74$\pm$0.19 (29) &   5.28$\pm$0.24 (9) &   5.15$\pm$0.16 (1) &   5.08$\pm$0.20 (8) &   5.41$\pm$0.18 (2) &  5.05$\pm$0.22 (16) & 4.99$\pm$0.26 (63) \\
   Fe& 7.47$\pm$0.13 (95) &  7.53$\pm$0.08 (82) &  7.44$\pm$0.11 (17) & 7.93$\pm$0.13 (438) &  7.39$\pm$0.08 (62) &  7.19$\pm$0.13 (24) &  7.47$\pm$0.11 (87) &  6.63$\pm$0.18 (21) & 7.27$\pm$0.09 (154) &7.13$\pm$0.12 (652) \\
   Co & 6.02$\pm$0.17 (2) &   5.59$\pm$0.13 (2) &                 $-$ &   5.40$\pm$0.05 (3) &   6.36$\pm$0.15 (1) &                 $-$ &                 $-$ &                 $-$ &   5.02$\pm$0.17 (1) & 4.62$\pm$0.22 (27) \\
   Ni& 6.12$\pm$0.19 (37) &  6.49$\pm$0.11 (27) &   6.17$\pm$0.11 (2) & 6.96$\pm$0.13 (100) &  6.28$\pm$0.13 (27) &   5.85$\pm$0.09 (3) &  6.30$\pm$0.15 (19) &   5.50$\pm$0.08 (7) &  6.08$\pm$0.16 (49) &5.86$\pm$0.15 (157) \\
   Cu & 3.88$\pm$0.37 (3) &   4.17$\pm$0.06 (3) &                 $-$ &   4.92$\pm$0.16 (2) &   4.42$\pm$0.15 (2) &                 $-$ &                 $-$ &                 $-$ &   4.08$\pm$0.17 (2) &  3.73$\pm$0.11 (5) \\
   Zn & 4.16$\pm$0.17 (2) &   4.83$\pm$0.13 (2) &                 $-$ &   5.28$\pm$0.04 (3) &   4.53$\pm$0.15 (2) &                 $-$ &   4.57$\pm$0.16 (1) &   3.70$\pm$0.18 (1) &   4.57$\pm$0.17 (2) &  4.18$\pm$0.04 (4) \\
   Ga               & $-$ &                 $-$ &                 $-$ &   3.38$\pm$0.16 (1) &                 $-$ &                 $-$ &                 $-$ &                 $-$ &   2.96$\pm$0.17 (1) &  2.40$\pm$0.22 (1) \\
   Sr & 3.41$\pm$0.17 (1) &   4.00$\pm$0.13 (2) &                 $-$ &   4.00$\pm$0.16 (2) &   3.56$\pm$0.15 (2) &   0.79$\pm$0.16 (1) &   3.91$\pm$0.16 (2) &   0.61$\pm$0.18 (1) &   3.69$\pm$0.17 (2) &  2.66$\pm$0.11 (3) \\
    Y & 1.97$\pm$0.21 (4) &  3.01$\pm$0.19 (11) &                 $-$ &  3.04$\pm$0.11 (17) &   2.64$\pm$0.10 (7) &                 $-$ &   2.62$\pm$0.24 (4) &   2.20$\pm$0.18 (2) &  2.53$\pm$0.23 (16) & 1.99$\pm$0.25 (18) \\
   Zr & 3.31$\pm$0.17 (2) &   3.19$\pm$0.20 (6) &                 $-$ &   3.00$\pm$0.16 (3) &   3.06$\pm$0.22 (6) &                 $-$ &   3.64$\pm$0.16 (2) &                 $-$ &   3.01$\pm$0.11 (9) & 2.45$\pm$0.26 (24) \\
   Ba & 2.66$\pm$0.10 (3) &   3.60$\pm$0.14 (3) &   3.60$\pm$0.11 (1) &   3.80$\pm$0.16 (5) &   3.04$\pm$0.15 (2) &   1.88$\pm$0.16 (2) &   3.58$\pm$0.19 (3) &   1.32$\pm$0.18 (1) &   2.79$\pm$0.21 (3) &  2.31$\pm$0.09 (5) \\
   La & 0.60$\pm$0.17 (1) &   1.60$\pm$0.22 (3) &                 $-$ &   2.18$\pm$0.16 (2) &   1.32$\pm$0.15 (1) &                 $-$ &                 $-$ &                 $-$ &   1.63$\pm$0.22 (3) & 0.96$\pm$0.19 (13) \\
   Ce & 1.94$\pm$0.17 (2) &   2.40$\pm$0.13 (1) &                 $-$ &  2.79$\pm$0.03 (10) &                 $-$ &                 $-$ &                 $-$ &                 $-$ &   2.22$\pm$0.17 (2) & 1.60$\pm$0.20 (34) \\
   Pr & 0.65$\pm$0.17 (1) &                 $-$ &                 $-$ &   1.67$\pm$0.16 (2) &                 $-$ &                 $-$ &                 $-$ &                 $-$ &                 $-$ &  0.46$\pm$0.34 (7) \\
   Nd & 1.38$\pm$0.17 (1) &   2.10$\pm$0.10 (6) &                 $-$ &  2.40$\pm$0.10 (10) &   1.84$\pm$0.15 (2) &                 $-$ &   2.42$\pm$0.16 (1) &                 $-$ &   2.04$\pm$0.18 (4) & 1.29$\pm$0.23 (35) \\
   Sm               & $-$ &                 $-$ &                 $-$ &                 $-$ &                 $-$ &                 $-$ &                 $-$ &                 $-$ &                 $-$ &  0.93$\pm$0.22 (2) \\
   Eu               & $-$ &                 $-$ &                 $-$ &   2.01$\pm$0.16 (1) &                 $-$ &                 $-$ &                 $-$ &                 $-$ &                 $-$ &  2.08$\pm$0.22 (2) \\
   Gd               & $-$ &                 $-$ &                 $-$ &                 $-$ &                 $-$ &                 $-$ &                 $-$ &                 $-$ &                 $-$ &  1.03$\pm$0.17 (3) \\
   Dy               & $-$ &                 $-$ &                 $-$ &                 $-$ &                 $-$ &                 $-$ &                 $-$ &                 $-$ &                 $-$ &  0.58$\pm$0.28 (4) \\
   Er               & $-$ &                 $-$ &                 $-$ &                 $-$ &                 $-$ &                 $-$ &                 $-$ &                 $-$ &                 $-$ &  1.43$\pm$0.22 (2) \\

\bottomrule
\end{tabular}
\end{scriptsize}
\end{table}
\end{landscape}

\newpage

\setcounter{table}{0}
\begin{landscape}

\begin{table}
\begin{scriptsize}
\caption{continuation}
\begin{tabular}{lllllllllll}
\midrule
    & KIC\,6443122 &      KIC\,6670742 &      KIC\,7119530 &      KIC\,7122746 &      KIC\,7668791 &      KIC\,7748238 &      KIC\,7767565 &      KIC\,7773133 &      KIC\,7798339 &      KIC\,8211500\\
    C & 8.39$\pm$0.23 (4) &   8.58$\pm$0.05 (3) &   8.18$\pm$0.15 (3) &   8.43$\pm$0.07 (3) &  8.29$\pm$0.20 (13) &   8.45$\pm$0.15 (6) &   8.21$\pm$0.23 (5) &   8.78$\pm$0.11 (8) &  8.36$\pm$0.25 (27) &  8.51$\pm$0.11 (4) \\
    N               & $-$ &                 $-$ &                 $-$ &                 $-$ &   7.99$\pm$0.16 (1) &                 $-$ &                 $-$ &                 $-$ &   8.17$\pm$0.19 (1) &                $-$ \\
    O & 8.99$\pm$0.16 (1) &   9.11$\pm$0.16 (2) &   8.66$\pm$0.22 (2) &   8.86$\pm$0.17 (2) &   8.92$\pm$0.16 (2) &   8.78$\pm$0.16 (1) &   8.53$\pm$0.17 (2) &   8.94$\pm$0.16 (1) &   8.58$\pm$0.19 (2) &  8.82$\pm$0.18 (2) \\
   Ne               & $-$ &                 $-$ &                 $-$ &                 $-$ &                 $-$ &                 $-$ &                 $-$ &                 $-$ &                 $-$ &                $-$ \\
   Na & 7.04$\pm$0.16 (2) &                 $-$ &   6.77$\pm$0.22 (1) &   5.85$\pm$0.17 (1) &   6.79$\pm$0.10 (4) &   6.65$\pm$0.16 (2) &   6.42$\pm$0.17 (2) &   6.51$\pm$0.15 (4) &   5.75$\pm$0.12 (7) &  6.46$\pm$0.18 (2) \\
   Mg & 7.37$\pm$0.08 (4) &   7.86$\pm$0.19 (5) &   7.71$\pm$0.17 (5) &   7.76$\pm$0.16 (7) &   7.82$\pm$0.07 (9) &   7.87$\pm$0.16 (5) &   8.06$\pm$0.24 (8) &   7.70$\pm$0.09 (9) &  7.45$\pm$0.18 (10) &  7.95$\pm$0.18 (5) \\
   Al               & $-$ &                 $-$ &                 $-$ &                 $-$ &   6.77$\pm$0.16 (1) &                 $-$ &                 $-$ &                 $-$ &   6.25$\pm$0.19 (2) &                $-$ \\
   Si& 7.40$\pm$0.24 (14) &   7.34$\pm$0.27 (3) &   7.86$\pm$0.29 (6) &   7.31$\pm$0.40 (7) &  7.56$\pm$0.19 (15) &  7.74$\pm$0.22 (10) &  7.77$\pm$0.21 (17) &  7.62$\pm$0.24 (22) &  7.28$\pm$0.21 (36) &  7.36$\pm$0.18 (7) \\
    P               & $-$ &                 $-$ &                 $-$ &                 $-$ &                 $-$ &                 $-$ &                 $-$ &                 $-$ &                 $-$ &                $-$ \\
    S & 7.39$\pm$0.11 (3) &                 $-$ &                 $-$ &   6.92$\pm$0.17 (1) &   7.46$\pm$0.26 (5) &   7.59$\pm$0.16 (2) &   7.27$\pm$0.17 (2) &   7.47$\pm$0.02 (3) &   7.06$\pm$0.13 (9) &  7.63$\pm$0.18 (2) \\
   Cl               & $-$ &                 $-$ &                 $-$ &                 $-$ &                 $-$ &                 $-$ &                 $-$ &                 $-$ &                 $-$ &                $-$ \\
    K               & $-$ &                 $-$ &                 $-$ &                 $-$ &                 $-$ &                 $-$ &                 $-$ &                 $-$ &                 $-$ &                $-$ \\
   Ca& 6.25$\pm$0.25 (12) &   6.17$\pm$0.11 (7) &   6.44$\pm$0.12 (6) &  6.32$\pm$0.17 (12) &  6.52$\pm$0.16 (29) &  6.65$\pm$0.14 (13) &  6.51$\pm$0.20 (23) &  6.62$\pm$0.19 (24) &  6.40$\pm$0.18 (41) & 6.18$\pm$0.23 (13) \\
   Sc & 2.86$\pm$0.12 (6) &   3.32$\pm$0.29 (4) &   3.29$\pm$0.30 (6) &   3.26$\pm$0.15 (5) &  3.23$\pm$0.17 (10) &   3.18$\pm$0.19 (6) &   3.30$\pm$0.09 (8) &  3.40$\pm$0.24 (11) &  3.10$\pm$0.19 (16) &  2.99$\pm$0.23 (5) \\
   Ti& 4.86$\pm$0.12 (17) &   4.86$\pm$0.15 (8) &   5.08$\pm$0.10 (8) &  5.13$\pm$0.12 (18) &  5.19$\pm$0.13 (47) &  5.01$\pm$0.16 (20) &  5.32$\pm$0.21 (31) &  5.32$\pm$0.19 (55) & 4.93$\pm$0.18 (115) & 5.13$\pm$0.20 (20) \\
    V & 4.61$\pm$0.16 (2) &                 $-$ &                 $-$ &                 $-$ &   4.41$\pm$0.14 (5) &   4.79$\pm$0.16 (2) &   4.96$\pm$0.15 (4) &   4.41$\pm$0.27 (5) &  4.02$\pm$0.24 (18) &  4.04$\pm$0.18 (1) \\
   Cr& 5.65$\pm$0.17 (16) &   5.63$\pm$0.14 (5) &   5.68$\pm$0.23 (5) &  5.58$\pm$0.14 (10) &  5.79$\pm$0.18 (59) &  5.55$\pm$0.16 (15) &  6.00$\pm$0.14 (31) &  5.92$\pm$0.16 (49) & 5.41$\pm$0.23 (125) & 5.88$\pm$0.14 (11) \\
   Mn & 5.21$\pm$0.28 (7) &   5.65$\pm$0.16 (2) &   5.71$\pm$0.22 (2) &   5.60$\pm$0.17 (2) &  5.29$\pm$0.17 (11) &   5.66$\pm$0.20 (5) &   5.23$\pm$0.14 (8) &  5.30$\pm$0.25 (15) &  5.01$\pm$0.25 (35) &  5.61$\pm$0.14 (3) \\
   Fe& 7.50$\pm$0.08 (45) &  7.31$\pm$0.11 (16) &  7.38$\pm$0.21 (19) &  7.45$\pm$0.12 (37) & 7.46$\pm$0.10 (117) &  7.49$\pm$0.11 (48) &  7.75$\pm$0.13 (98) & 7.64$\pm$0.11 (143) & 7.13$\pm$0.12 (352) & 7.56$\pm$0.09 (39) \\
   Co               & $-$ &                 $-$ &                 $-$ &                 $-$ &                 $-$ &                 $-$ &                 $-$ &   6.32$\pm$0.16 (2) &  4.81$\pm$0.30 (16) &                $-$ \\
   Ni& 6.47$\pm$0.22 (23) &   5.93$\pm$0.13 (3) &   6.80$\pm$0.31 (4) &   6.21$\pm$0.28 (9) &  6.18$\pm$0.19 (34) &  6.43$\pm$0.19 (18) &  6.63$\pm$0.14 (23) &  6.22$\pm$0.15 (47) & 5.87$\pm$0.17 (100) & 6.51$\pm$0.22 (11) \\
   Cu & 4.51$\pm$0.16 (1) &                 $-$ &                 $-$ &                 $-$ &   4.19$\pm$0.16 (2) &   4.35$\pm$0.16 (2) &   4.43$\pm$0.17 (1) &   4.29$\pm$0.16 (2) &   3.35$\pm$0.19 (1) &                $-$ \\
   Zn & 4.60$\pm$0.16 (2) &                 $-$ &                 $-$ &   4.39$\pm$0.17 (1) &   4.45$\pm$0.16 (1) &                 $-$ &   4.76$\pm$0.17 (2) &   4.29$\pm$0.16 (2) &   3.97$\pm$0.19 (2) &  4.48$\pm$0.18 (1) \\
   Ga               & $-$ &                 $-$ &                 $-$ &                 $-$ &                 $-$ &                 $-$ &                 $-$ &                 $-$ &                 $-$ &                $-$ \\
   Sr & 3.43$\pm$0.16 (1) &   1.23$\pm$0.16 (1) &   1.48$\pm$0.22 (1) &   2.90$\pm$0.17 (1) &   3.93$\pm$0.16 (1) &   3.27$\pm$0.16 (1) &   4.10$\pm$0.17 (2) &   3.74$\pm$0.16 (2) &   3.20$\pm$0.04 (3) &  3.02$\pm$0.18 (1) \\
    Y & 2.70$\pm$0.09 (8) &                 $-$ &   2.68$\pm$0.33 (3) &   2.05$\pm$0.17 (2) &  2.55$\pm$0.19 (10) &   2.51$\pm$0.24 (4) &   2.95$\pm$0.24 (6) &   2.61$\pm$0.20 (8) &  2.30$\pm$0.16 (19) &  2.57$\pm$0.28 (3) \\
   Zr & 2.90$\pm$0.16 (2) &                 $-$ &                 $-$ &   3.18$\pm$0.17 (1) &  2.78$\pm$0.21 (10) &   3.63$\pm$0.04 (3) &   3.14$\pm$0.17 (4) &   3.09$\pm$0.20 (7) &  2.92$\pm$0.13 (19) &  3.45$\pm$0.18 (1) \\
   Ba & 3.13$\pm$0.16 (2) &   3.01$\pm$0.16 (1) &   2.64$\pm$0.22 (2) &   2.29$\pm$0.17 (2) &   2.53$\pm$0.16 (2) &   2.61$\pm$0.16 (2) &   3.85$\pm$0.09 (4) &   3.10$\pm$0.16 (2) &   2.89$\pm$0.23 (4) &  3.37$\pm$0.18 (2) \\
   La & 1.52$\pm$0.16 (1) &                 $-$ &                 $-$ &                 $-$ &   1.49$\pm$0.16 (2) &                 $-$ &   2.45$\pm$0.17 (2) &   1.37$\pm$0.16 (2) &  1.29$\pm$0.19 (11) &                $-$ \\
   Ce               & $-$ &                 $-$ &                 $-$ &                 $-$ &   2.27$\pm$0.22 (4) &                 $-$ &   2.63$\pm$0.17 (1) &   2.02$\pm$0.16 (1) &  1.69$\pm$0.22 (34) &                $-$ \\
   Pr               & $-$ &                 $-$ &                 $-$ &                 $-$ &                 $-$ &                 $-$ &                 $-$ &                 $-$ &   1.08$\pm$0.19 (2) &                $-$ \\
   Nd               & $-$ &                 $-$ &                 $-$ &                 $-$ &   2.38$\pm$0.16 (2) &                 $-$ &   2.60$\pm$0.17 (2) &   1.65$\pm$0.14 (3) &  1.44$\pm$0.26 (31) &                $-$ \\
   Sm               & $-$ &                 $-$ &                 $-$ &                 $-$ &                 $-$ &                 $-$ &                 $-$ &                 $-$ &   1.43$\pm$0.19 (2) &                $-$ \\
   Eu               & $-$ &                 $-$ &                 $-$ &                 $-$ &                 $-$ &                 $-$ &                 $-$ &                 $-$ &   0.30$\pm$0.19 (2) &                $-$ \\
   Gd               & $-$ &                 $-$ &                 $-$ &                 $-$ &                 $-$ &                 $-$ &                 $-$ &                 $-$ &   1.19$\pm$0.08 (3) &                $-$ \\
   Dy               & $-$ &                 $-$ &                 $-$ &                 $-$ &                 $-$ &                 $-$ &                 $-$ &                 $-$ &   1.37$\pm$0.04 (4) &                $-$ \\
   Er               & $-$ &                 $-$ &                 $-$ &                 $-$ &                 $-$ &                 $-$ &                 $-$ &                 $-$ &   1.38$\pm$0.19 (1) &                $-$ \\

\bottomrule
\end{tabular}
\end{scriptsize}
\end{table}
\end{landscape}

\newpage

\setcounter{table}{0}
\begin{landscape}

\begin{table}
\begin{scriptsize}
\caption{continuation}
\begin{tabular}{lllllllllll}
\midrule
    & KIC\,8222685 &      KIC\,8694723 &      KIC\,8827821 &      KIC\,8881697 &      KIC\,9229318 &      KIC\,9349245 &      KIC\,9408694 &      KIC\,9410862 &      KIC\,9650390 &      KIC\,9656348\\
    C & 8.07$\pm$0.11 (5) &   8.19$\pm$0.06 (9) &   8.15$\pm$0.10 (3) &   8.72$\pm$0.25 (3) &   8.60$\pm$0.24 (9) &  8.38$\pm$0.16 (13) &  8.59$\pm$0.19 (20) &  8.33$\pm$0.22 (10) &   8.61$\pm$0.15 (1) & 8.31$\pm$0.26 (15) \\
    N               & $-$ &                 $-$ &                 $-$ &                 $-$ &                 $-$ &   8.40$\pm$0.15 (1) &                 $-$ &                 $-$ &                 $-$ &                $-$ \\
    O & 8.92$\pm$0.14 (1) &                 $-$ &   8.66$\pm$0.15 (1) &   9.14$\pm$0.19 (2) &   9.12$\pm$0.16 (2) &   8.78$\pm$0.07 (4) &   9.08$\pm$0.24 (1) &                 $-$ &   8.95$\pm$0.15 (1) &  8.68$\pm$0.23 (2) \\
   Ne               & $-$ &                 $-$ &                 $-$ &                 $-$ &                 $-$ &                 $-$ &                 $-$ &                 $-$ &                 $-$ &                $-$ \\
   Na & 6.51$\pm$0.14 (1) &   5.85$\pm$0.11 (7) &   6.32$\pm$0.15 (2) &                 $-$ &   6.95$\pm$0.21 (4) &   7.18$\pm$0.21 (6) &   6.52$\pm$0.17 (9) &   6.03$\pm$0.11 (9) &                 $-$ &  6.20$\pm$0.28 (3) \\
   Mg & 7.50$\pm$0.08 (5) &  7.22$\pm$0.12 (11) &   7.99$\pm$0.15 (7) &   7.95$\pm$0.14 (6) &   7.92$\pm$0.14 (8) &   8.11$\pm$0.12 (9) &  7.82$\pm$0.17 (13) &   7.54$\pm$0.05 (8) &   8.07$\pm$0.15 (2) &  7.51$\pm$0.19 (6) \\
   Al               & $-$ &                 $-$ &                 $-$ &                 $-$ &   6.67$\pm$0.16 (1) &                 $-$ &                 $-$ &   5.78$\pm$0.17 (1) &                 $-$ &                $-$ \\
   Si& 7.47$\pm$0.25 (11) &  7.24$\pm$0.28 (34) &  7.64$\pm$0.23 (12) &   7.53$\pm$0.25 (5) &  7.73$\pm$0.18 (16) &  7.93$\pm$0.20 (23) &  7.67$\pm$0.29 (44) &  7.32$\pm$0.19 (46) &   7.29$\pm$0.19 (3) & 7.03$\pm$0.25 (10) \\
    P               & $-$ &                 $-$ &                 $-$ &                 $-$ &                 $-$ &                 $-$ &                 $-$ &                 $-$ &                 $-$ &                $-$ \\
    S & 7.14$\pm$0.14 (2) &   7.00$\pm$0.17 (2) &   7.55$\pm$0.15 (2) &   7.83$\pm$0.19 (1) &   7.57$\pm$0.11 (3) &   7.50$\pm$0.11 (7) &  7.47$\pm$0.16 (11) &   7.06$\pm$0.17 (2) &                 $-$ &  6.58$\pm$0.23 (1) \\
   Cl               & $-$ &                 $-$ &                 $-$ &                 $-$ &                 $-$ &                 $-$ &                 $-$ &                 $-$ &                 $-$ &                $-$ \\
    K               & $-$ &                 $-$ &                 $-$ &                 $-$ &                 $-$ &                 $-$ &                 $-$ &                 $-$ &                 $-$ &                $-$ \\
   Ca& 6.27$\pm$0.13 (12) &  5.99$\pm$0.16 (47) &  6.59$\pm$0.20 (15) &   6.50$\pm$0.20 (7) &  6.67$\pm$0.18 (17) &  6.87$\pm$0.18 (22) &  6.58$\pm$0.19 (38) &  6.22$\pm$0.18 (33) &   6.56$\pm$0.14 (4) & 5.85$\pm$0.12 (16) \\
   Sc & 2.85$\pm$0.08 (5) &  2.85$\pm$0.10 (23) &   3.43$\pm$0.13 (6) &   3.16$\pm$0.20 (5) &   3.55$\pm$0.04 (6) &   3.38$\pm$0.13 (7) &  3.62$\pm$0.13 (13) &  2.88$\pm$0.10 (16) &   3.77$\pm$0.24 (5) &  2.38$\pm$0.25 (5) \\
   Ti& 4.77$\pm$0.09 (19) & 4.65$\pm$0.16 (111) &  5.11$\pm$0.12 (16) &  5.07$\pm$0.23 (15) &  5.42$\pm$0.13 (25) &  5.30$\pm$0.17 (33) &  5.25$\pm$0.20 (89) & 4.74$\pm$0.15 (117) &   4.88$\pm$0.09 (5) & 4.40$\pm$0.20 (34) \\
    V & 4.61$\pm$0.14 (1) &  3.28$\pm$0.26 (11) &   4.92$\pm$0.15 (2) &   4.07$\pm$0.19 (1) &   4.85$\pm$0.25 (5) &   4.81$\pm$0.15 (5) &  4.63$\pm$0.23 (13) &  3.57$\pm$0.15 (22) &                 $-$ &  4.07$\pm$0.22 (5) \\
   Cr& 5.52$\pm$0.14 (13) & 5.18$\pm$0.17 (105) &  5.69$\pm$0.22 (15) &   5.78$\pm$0.18 (8) &  5.96$\pm$0.17 (25) &  5.97$\pm$0.14 (33) &  5.79$\pm$0.22 (97) & 5.30$\pm$0.17 (110) &   5.98$\pm$0.13 (5) & 5.01$\pm$0.20 (27) \\
   Mn & 5.14$\pm$0.15 (4) &  4.94$\pm$0.19 (32) &   5.43$\pm$0.15 (3) &   5.67$\pm$0.19 (2) &   5.75$\pm$0.26 (6) &   5.74$\pm$0.10 (6) &  5.25$\pm$0.24 (29) &  4.97$\pm$0.19 (32) &   6.13$\pm$0.15 (1) &  4.65$\pm$0.34 (7) \\
   Fe& 7.20$\pm$0.07 (47) & 7.02$\pm$0.10 (431) &  7.60$\pm$0.07 (34) &  7.49$\pm$0.11 (26) &  7.75$\pm$0.08 (53) &  7.92$\pm$0.11 (89) & 7.53$\pm$0.13 (338) & 7.19$\pm$0.13 (427) &  7.59$\pm$0.11 (13) & 6.84$\pm$0.11 (81) \\
   Co               & $-$ &   4.60$\pm$0.12 (7) &                 $-$ &                 $-$ &   5.41$\pm$0.16 (1) &   6.40$\pm$0.15 (2) &  5.04$\pm$0.22 (11) &  4.70$\pm$0.06 (15) &                 $-$ &  4.73$\pm$0.23 (2) \\
   Ni& 6.05$\pm$0.18 (15) &  5.76$\pm$0.12 (91) &  6.47$\pm$0.17 (14) &   6.42$\pm$0.19 (6) &  6.61$\pm$0.14 (31) &  6.89$\pm$0.16 (45) &  6.28$\pm$0.15 (87) & 5.93$\pm$0.15 (138) &   6.66$\pm$0.15 (2) & 5.86$\pm$0.37 (27) \\
   Cu & 4.23$\pm$0.14 (2) &   3.66$\pm$0.10 (3) &   4.80$\pm$0.15 (2) &                 $-$ &   5.24$\pm$0.16 (2) &   4.62$\pm$0.15 (2) &   4.10$\pm$0.11 (4) &   3.78$\pm$0.01 (3) &                 $-$ &  4.13$\pm$0.23 (2) \\
   Zn & 4.47$\pm$0.14 (2) &   4.24$\pm$0.17 (2) &   4.40$\pm$0.15 (1) &   4.66$\pm$0.19 (1) &   4.46$\pm$0.16 (1) &   5.18$\pm$0.15 (2) &   4.61$\pm$0.24 (2) &   4.18$\pm$0.08 (3) &                 $-$ &  3.98$\pm$0.23 (2) \\
   Ga               & $-$ &   2.48$\pm$0.17 (1) &                 $-$ &                 $-$ &                 $-$ &                 $-$ &   3.91$\pm$0.24 (2) &                 $-$ &                 $-$ &                $-$ \\
   Sr & 3.01$\pm$0.14 (1) &   2.72$\pm$0.17 (2) &   3.66$\pm$0.15 (1) &   3.00$\pm$0.19 (1) &   4.00$\pm$0.16 (1) &   3.40$\pm$0.15 (1) &   3.63$\pm$0.25 (3) &   3.24$\pm$0.17 (1) &   2.42$\pm$0.15 (1) &  2.96$\pm$0.23 (2) \\
    Y & 2.22$\pm$0.31 (3) &  1.75$\pm$0.14 (14) &   2.63$\pm$0.21 (3) &   1.63$\pm$0.19 (1) &   2.89$\pm$0.21 (7) &   2.70$\pm$0.18 (7) &  2.42$\pm$0.14 (11) &  1.65$\pm$0.15 (16) &   1.90$\pm$0.15 (1) &  1.71$\pm$0.23 (7) \\
   Zr & 2.73$\pm$0.14 (2) &  2.49$\pm$0.28 (19) &   3.04$\pm$0.14 (3) &   3.12$\pm$0.19 (1) &   3.22$\pm$0.08 (4) &   3.13$\pm$0.33 (4) &  2.95$\pm$0.20 (14) &   2.75$\pm$0.34 (6) &                 $-$ &  2.58$\pm$0.08 (5) \\
   Ba & 2.17$\pm$0.14 (2) &   2.22$\pm$0.15 (4) &   3.14$\pm$0.15 (2) &   2.76$\pm$0.19 (2) &   3.08$\pm$0.16 (2) &   3.62$\pm$0.14 (5) &   2.76$\pm$0.16 (5) &   2.13$\pm$0.25 (4) &                 $-$ &  1.65$\pm$0.35 (3) \\
   La & 0.77$\pm$0.14 (1) &   0.89$\pm$0.24 (4) &   1.74$\pm$0.15 (1) &                 $-$ &   2.16$\pm$0.16 (2) &   2.57$\pm$0.11 (3) &   1.48$\pm$0.14 (8) &   0.96$\pm$0.27 (5) &                 $-$ &  1.16$\pm$0.23 (2) \\
   Ce               & $-$ &  1.44$\pm$0.14 (16) &                 $-$ &                 $-$ &   2.25$\pm$0.16 (2) &   2.59$\pm$0.10 (4) &  1.85$\pm$0.21 (15) &  1.44$\pm$0.14 (10) &                 $-$ &  1.47$\pm$0.26 (3) \\
   Pr               & $-$ &   0.20$\pm$0.17 (2) &                 $-$ &                 $-$ &                 $-$ &                 $-$ &   1.26$\pm$0.33 (3) &   0.72$\pm$0.17 (1) &                 $-$ &                $-$ \\
   Nd               & $-$ &  1.12$\pm$0.20 (12) &                 $-$ &                 $-$ &   2.21$\pm$0.16 (1) &   2.45$\pm$0.15 (2) &  1.72$\pm$0.26 (14) &  1.11$\pm$0.26 (18) &                 $-$ &  1.75$\pm$0.23 (2) \\
   Sm               & $-$ &   0.70$\pm$0.17 (1) &                 $-$ &                 $-$ &                 $-$ &                 $-$ &   1.25$\pm$0.24 (1) &   2.10$\pm$0.17 (1) &                 $-$ &                $-$ \\
   Eu               & $-$ &   0.68$\pm$0.17 (1) &                 $-$ &                 $-$ &                 $-$ &                 $-$ &                 $-$ &                 $-$ &                 $-$ &                $-$ \\
   Gd               & $-$ &   1.11$\pm$0.17 (1) &                 $-$ &                 $-$ &                 $-$ &                 $-$ &   1.77$\pm$0.24 (2) &                 $-$ &                 $-$ &                $-$ \\
   Dy               & $-$ &   0.80$\pm$0.17 (1) &                 $-$ &                 $-$ &                 $-$ &                 $-$ &   1.67$\pm$0.24 (2) &                 $-$ &                 $-$ &                $-$ \\
   Er               & $-$ &                 $-$ &                 $-$ &                 $-$ &                 $-$ &                 $-$ &                 $-$ &                 $-$ &                 $-$ &                $-$ \\

\bottomrule
\end{tabular}
\end{scriptsize}
\end{table}
\end{landscape}

\newpage

\setcounter{table}{0}
\begin{landscape}

\begin{table}
\begin{scriptsize}
\caption{continuation}
\begin{tabular}{lllllll}
\midrule
    & KIC\,9828226 &      KIC\,9845907 &      KIC\,9941662 &      KIC\,9970568              \\                                                                                                     
    C & 8.43$\pm$0.16 (2) &  7.90$\pm$0.20 (19) &   8.25$\pm$0.34 (4) &   7.86$\pm$0.24 (2) \\                                                                                                         
    N               & $-$ &   7.58$\pm$0.18 (1) &                 $-$ &                 $-$ \\                                                                                                                  
    O & 8.89$\pm$0.16 (1) &   8.57$\pm$0.04 (3) &   8.55$\pm$0.20 (1) &   9.26$\pm$0.24 (1) \\                                                                                               
   Ne               & $-$ &                 $-$ &                 $-$ &                 $-$ \\                                                                                                      
   Na               & $-$ &   6.08$\pm$0.15 (7) &   6.66$\pm$0.20 (1) &   6.66$\pm$0.24 (1) \\                                                                                                 
   Mg & 6.52$\pm$0.16 (2) &  7.47$\pm$0.13 (13) &   7.73$\pm$0.22 (4) &   7.74$\pm$0.27 (5) \\                                                                                               
   Al               & $-$ &   6.15$\pm$0.18 (1) &                 $-$ &                 $-$ \\                                                                                                 
   Si & 6.60$\pm$0.16 (2) &  7.34$\pm$0.16 (24) &   7.73$\pm$0.14 (6) &   6.89$\pm$0.34 (4) \\                                                                                               
    P               & $-$ &                 $-$ &                 $-$ &                 $-$ \\                    
    S               & $-$ &   7.02$\pm$0.20 (8) &   7.54$\pm$0.20 (1) &   7.94$\pm$0.24 (2) \\                                                                                    
   Cl               & $-$ &                 $-$ &                 $-$ &                 $-$ \\                                                           
    K               & $-$ &                 $-$ &                 $-$ &                 $-$ \\                                                                    
   Ca & 6.21$\pm$0.16 (2) &  6.01$\pm$0.11 (32) &   5.69$\pm$0.19 (8) &   6.37$\pm$0.21 (8) \\                                                              
   Sc & 3.24$\pm$0.16 (2) &   2.61$\pm$0.12 (9) &   2.94$\pm$0.54 (3) &   3.14$\pm$0.25 (3) \\                                                                
   Ti & 3.91$\pm$0.13 (4) &  4.79$\pm$0.15 (78) &  5.13$\pm$0.16 (16) &  5.18$\pm$0.25 (12) \\                                                             
    V               & $-$ &  4.27$\pm$0.21 (20) &   4.70$\pm$0.20 (2) &                 $-$ \\                                                               
   Cr & 4.71$\pm$0.16 (4) &  5.39$\pm$0.20 (84) &  6.07$\pm$0.19 (10) &   5.41$\pm$0.30 (8) \\                                                           
   Mn               & $-$ &  5.23$\pm$0.20 (32) &   5.20$\pm$0.13 (4) &   4.91$\pm$0.24 (2) \\                                                  
   Fe& 6.37$\pm$0.21 (11) & 7.10$\pm$0.11 (272) &  7.63$\pm$0.06 (39) &  7.34$\pm$0.09 (21) \\                           
   Co               & $-$ &   5.18$\pm$0.04 (3) &                 $-$ &                 $-$ \\                                                       
   Ni               & $-$ &  5.79$\pm$0.23 (62) &   6.49$\pm$0.11 (8) &   6.07$\pm$0.28 (6) \\                                             
   Cu               & $-$ &   3.93$\pm$0.18 (2) &                 $-$ &                 $-$ \\                                                                      
   Zn               & $-$ &                 $-$ &   4.54$\pm$0.20 (1) &   4.90$\pm$0.24 (1) \\                                                                  
   Ga               & $-$ &                 $-$ &                 $-$ &                 $-$ \\                                                                    
   Sr               & $-$ &   2.61$\pm$0.18 (2) &   3.89$\pm$0.20 (1) &   1.99$\pm$0.24 (1) \\                                                                  
    Y               & $-$ &   2.11$\pm$0.19 (8) &   2.94$\pm$0.20 (2) &   3.14$\pm$0.24 (2) \\                                                                          
   Zr               & $-$ &  2.57$\pm$0.19 (16) &   2.95$\pm$0.20 (1) &   2.93$\pm$0.24 (1) \\                                                                                
   Ba & 1.23$\pm$0.16 (1) &   2.01$\pm$0.14 (4) &   3.97$\pm$0.20 (2) &   3.47$\pm$0.24 (2) \\                                                                      
   La               & $-$ &   1.43$\pm$0.13 (3) &                 $-$ &                 $-$ \\                                                                           
   Ce               & $-$ &   2.05$\pm$0.17 (9) &                 $-$ &                 $-$ \\                                                                       
   Pr               & $-$ &                 $-$ &                 $-$ &                 $-$ \\                                                                        
   Nd               & $-$ &   1.44$\pm$0.28 (6) &                 $-$ &                 $-$ \\                                                                               
   Sm               & $-$ &                 $-$ &                 $-$ &                 $-$ \\                                                                            
   Eu               & $-$ &   0.87$\pm$0.18 (2) &                 $-$ &                 $-$ \\                                                                                   
   Gd               & $-$ &                 $-$ &                 $-$ &                 $-$ \\                                                                       
   Dy               & $-$ &                 $-$ &                 $-$ &                 $-$ \\                                                                     
   Er               & $-$ &   1.65$\pm$0.18 (1) &                 $-$ &                 $-$ \\                     

\bottomrule
\end{tabular}
\end{scriptsize}
\end{table}
\end{landscape}

\newpage

%%%%%%%%%%%%%%%%%%%%%%%%%%%%%%%%%%%%%%%%%%%%%%%%%%%%%%%%%%%%%%%%%%%%%%%%%%%%%%%%%%%%%%%%%%%%%%%%%%%%%%%%%%%%%%%%%%%%%

\label{lastpage}

\end{document}